\documentclass[a4paper,11pt,fleqn]{article}

\usepackage[ansinew]{inputenc}
\usepackage{hyperref}
\usepackage[mathscr]{eucal}
\usepackage{graphicx}
\usepackage[font=footnotesize,labelsep=period,justification=justified,figureposition=bottom,tableposition=top]{caption}
\usepackage{subcaption}
\usepackage{multirow}
\usepackage{amsmath,amssymb,amsthm}
\usepackage{comment}
\allowdisplaybreaks

\hypersetup{colorlinks, linkcolor=blue, citecolor=blue, urlcolor=blue}

\newcommand{\todo}[1][\null]{\ensuremath{\clubsuit}}

\newcommand{\noprint}[1]{}

{\theoremstyle{definition}

\newtheorem*{remark*}{Remark}
}

\begin{document}

\par\noindent {\LARGE\bf
Physics-informed neural networks\\ for the shallow-water equations on the sphere
\par}

\vspace{4mm}\par\noindent {\large
Alex Bihlo$^\dag$ and Roman O.\ Popovych$^{\ddag}$
\par}

\vspace{4mm}\par\noindent{\it
$^{\dag}$Department of Mathematics and Statistics, Memorial University of Newfoundland,\\
$\phantom{^{\dag}}$~St.\ John's (NL), A1C 5S7, Canada
}

\vspace{2mm}\par\noindent{\it
$^\ddag$\,Fakult\"at f\"ur Mathematik, Universit\"at Wien, Oskar-Morgenstern-Platz 1, A-1090 Wien, Austria%
\\
$\phantom{^\ddag}$Institute of Mathematics of NAS of Ukraine, 3 Tereshchenkivska Str., 01024 Kyiv, Ukraine
}

\vspace{2mm}\par\noindent {\it
\textup{E-mails:} abihlo@mun.ca,
rop@imath.kiev.ua
}\par

\vspace{12mm}\par\noindent\hspace*{10mm}\parbox{140mm}{\small
We propose the use of physics-informed neural networks for solving the shallow-water equations on the sphere in the meteorological context. Physics-informed neural networks are trained to satisfy the differential equations along with the prescribed initial and boundary data, and thus can be seen as an alternative approach to solving differential equations compared to traditional numerical approaches such as finite difference, finite volume or spectral methods. We discuss the training difficulties of physics-informed neural networks for the shallow-water equations on the sphere and propose a simple multi-model approach to tackle test cases of comparatively long time intervals. Here we train a sequence of neural networks instead of a single neural network for the entire integration interval. We also avoid the use of a boundary value loss by encoding the boundary conditions in a custom neural network layer. We illustrate the abilities of the method by solving the most prominent test cases proposed by Williamson et al.\ [\textit{J. Comput. Phys.} \textbf{102} (1992), 211--224].
\par}\vspace{7mm}

\noprint{
MSC: 68T07, 76U60, 86A10
68-XX  Computer science {For papers containing software, source code, etc. in a specific mathematical area, see the classification number -04 in that area}
 68Txx	Artificial intelligence
   68T07  Artificial neural networks and deep learning
76-XX  Fluid mechanics {For general continuum mechanics, see 74Axx, or other parts of 74-XX}
  76Uxx Rotating fluids
   76U60  Geophysical flows [See also 86A05, 86A10]
86-XX  Geophysics [See also 76U05, 76V05]
  86Axx Geophysics [See also 76U05, 76V05]
   86A10  Meteorology and atmospheric physics [See also 76Bxx, 76E20, 76N15, 76Q05, 76Rxx, 76U05]
}

\section{Introduction}\label{sec:IntroductionSWEDeepLearning}

Deep learning has seen an unprecedented rise in popularity over the last 10 years or so. Propelled by the seminal contribution~\cite{kriz12a}, which won the 2012 ImageNet competition by a large margin, it was shown that deep artificial neural networks can be trained successfully. Since then, deep neural networks have achieved state-of-the art in many domains such as image classification and computer vision~\cite{huan17a}, natural language processing~\cite{brow20a} and reinforcement learning~\cite{mnih13a,silv17a}.

The rise in popularity of deep learning has also readily reached geophysical fluid dynamics. Many groups around the world have successfully used deep neural networks for weather forecasting and weather analysis related tasks. For example, in~\cite{sche19a,weyn19a,weyn20a} convolutional and recurrent neural networks have been used to predict various meteorological fields, such as the 500 hPa geopotential height, with these networks being trained against several decades of reanalysis data. The problem of ensemble-prediction was considered in~\cite{bihl21a,sche20a} using generative adversarial networks with Monte-Carlo dropout as well as using re-training of neural networks. Weather nowcasting with convolutional LSTM networks was proposed e.g.\ in~\cite{bihl19a,xing15a}. Down-scaling of climate data with neural networks was put forth in~\cite{moua17a}. There is also active research on using deep learning for sub-grid scale parameterizations, see e.g.~\cite{gent18a}.

While the above mentioned works demonstrate impressively the potential of deep learning in meteorology, they largely ignore that meteorological data, as in other fields of natural sciences, evolves based on well-known systems of differential equations~\cite{holt04a,sato04a}. As such, it is a natural question to ask whether one should ignore these differential equations altogether in favor of a purely data-driven approach as was done in the above approaches. This issue is receiving increased attention in the meteorological community, with various approaches being proposed to include physical domain knowledge to machine learning, see for example~\cite{kash21a} for a recent high-level overview of activities in this direction.

Recently, in~\cite{rais19a} a new framework for solving differential equations based on deep neural networks was proposed, which is referred to as \textit{physics-informed neural networks} (PINNs). The idea is to train neural networks not solely based on data, as is traditionally done in the field of deep learning, but to also enforce the underlying system of differential equations over a domain~$\Omega$. Practically, this is done by defining a neural network that accepts as input the independent variables of the system
and outputs the dependent variables, which are known as the prognostic variables in weather and climate models. This is akin to learning an interpolating function for the solution of an initial boundary-value problem for the system of differential equations, with the principal difference being that the interpolating function used is not a (piecewise) polynomial but a neural network. Training of this neural network is done by enforcing that (i) the system of differential equations, (ii) the initial condition and (iii) the boundary conditions hold on finitely many so-called collocation points, which are sampled (typically at random) over the respective domains of the independent variables. In this sense, the role of collocation points for physics-informed neural networks is analogous to that of the spatio-temporal computational mesh points for finite difference or finite element methods. The neural network then learns to satisfy these constraints by minimizing a suitable mean-squared loss function. Once trained over the collection of collocation points, again akin to standard interpolation, the neural network solution to the system of differential equations can be used to evaluate this solution at any point over the domain~$\Omega$.

In other words, physics-informed neural networks are a competing approach for solving differential equations that by-passes any numerical discretization of the system of differential equations at hand. Given the importance of the shallow-water equations as a test bed for new numerical methods proposed in the atmospheric sciences, see e.g.~\cite{bihl12b,brec19a,flye12a,ring02a,thub15a} for various ``traditional'' numerical methods in this regard, it is thus instructive to investigate whether physics-informed neural networks could provide a feasible approach to numerically solving the governing equations in geophysical fluid dynamics. We should also like to note that physics-informed neural networks are thus in rather stark contrast to the approaches used so far for deep learning in the atmospheric sciences, which were devoted to learning a mapping from given data of discrete spatio-temporal steps of meteorological fields to future discrete spatio-temporal steps of the same fields.

Physics-informed neural networks have since been developed for many systems of differential equations with applications in science and engineering, see e.g.~\cite{jagt20c,jagt20a,jin21a,kiss20a,mao20a,meng20a} for some recent contributions. The purpose of the present paper is to show their application as a potential alternative solver for the governing equations of fluid dynamics, by proposing physics-informed neural networks for the shallow-water equations on a rotating sphere. Note that while the shallow-water equations can be used to study atmospheric waves, surface water waves and also internal water waves, in the following we are solely interested in the meteorological context.

The remainder of this paper is organized as follows. In Section~\ref{sec:TheoryPINNsSWE} we present a review of the theory behind physics-informed neural networks. Section~\ref{sec:PINNsSWE} discusses the implementation of such physics-informed neural networks for the shallow-water equations in standard latitude--longitude coordinates, which in particular demonstrates that these networks can readily circumvent the pole problem, which plagues most discretizations of the shallow-water equations in spherical geometry. Section~\ref{sec:ResultsPINNsSWE} presents the numerical results of physics-informed neural networks for the shallow-water equations by considering the classical benchmark test suite proposed in~\cite{will92Ay}. The final Section~\ref{sec:ConclusionsPINNsSWE} contains a summary along with some discussions on potential future research avenues on physics-informed neural networks in meteorology.

\section{Physics-informed neural networks}\label{sec:TheoryPINNsSWE}

\subsection{Neural network architectures}

Densely connected neural networks, as are traditionally used for physics-informed neural networks, are among the simplest neural network architectures within the field of deep learning. Mathematically, these networks can be represented as nested nonlinear transformations of simple affine transformations, see e.g.~\cite{high19a} for a mathematical introduction to this field.

\begin{figure}[!ht]
\centering
\includegraphics[scale=0.9]{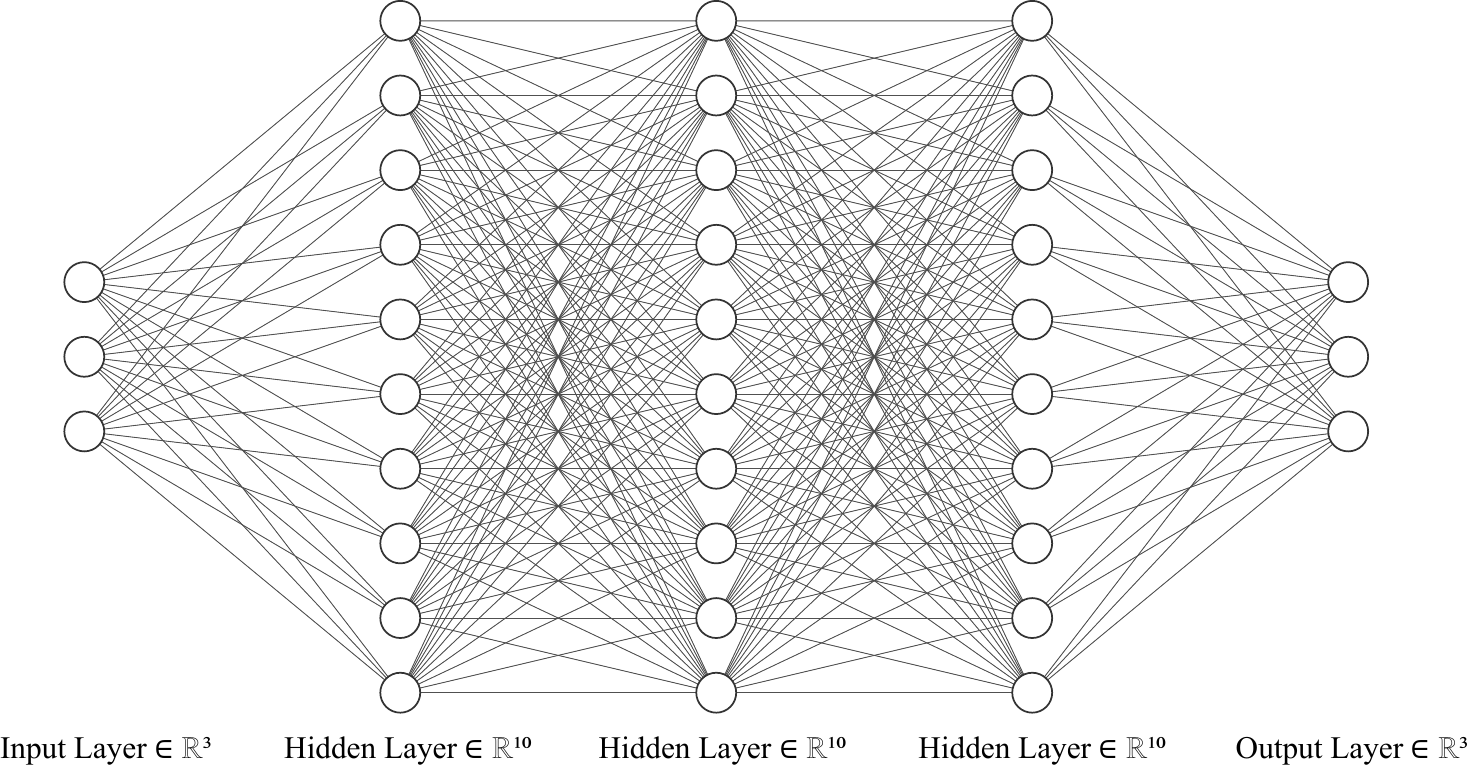}
\caption{A five layer densely connected neural network that has 3 hidden layers with 10 units per each layer, and input and output layers with 3 units each.}
\label{fig:DenselyConnectedNetwork}
\end{figure}

For example, the architecture given in Fig.~\ref{fig:DenselyConnectedNetwork} can be represented as a mapping from $\mathcal N^{\boldsymbol{\theta}}\colon \mathbb{R}^3\to\mathbb{R}^3$ of the form
\[
\mathbf{y} = \mathcal{N}^{\boldsymbol{\theta}}(\mathbf{x}) = \varphi_4(W_4\varphi_3(W_3\varphi_2 (W_2\varphi_1(W_1\mathbf{x}+\mathbf{b}_1)+\mathbf{b}_2)+\mathbf{b}_3)+\mathbf{b}_4),
\]
where $W_1\in\mathbb{R}^{10\times 3}$, $W_2,W_3\in\mathbb{R}^{10\times 10}$, $W_4\in\mathbb{R}^{3\times10}$ are the weight matrices and $\mathbf{b}_1,\mathbf{b}_2,\mathbf{b}_3\in\mathbb{R}^{10}$ and $\mathbf{b}_4\in\mathbb{R}^3$ are the associated bias vectors, collectively stored in the weight vector $\boldsymbol{\theta}$, that has to be learned by the network at training stage. The functions~$\varphi_i$, $i=1,\dots,4$ are some nonlinear activation functions, and are typically chosen as hyperbolic tangents, ReLUs, leaky ReLUs, or ELUs~\cite{gero19a}. These functions are assumed to act component-wise on tuples. Note that the wider (number of units per hidden layer) and deeper (number of hidden layers) neural networks become, the more parameters need to be learned. For example, the neural network depicted in Fig.~\ref{fig:DenselyConnectedNetwork} has a total of $(30+10)+(100+10)+(100+10)+(30+3) = 293$ weight and bias parameters that need to be tuned.

The success of physics-informed neural networks relies on the Universal Approximation Theorem~\cite{cybe89a} asserting that densely connected neural networks with suitable nonlinear activation functions can represent wide classes of well-behaved functions. As such, they are theoretically capable of representing the solutions of well-posed differential equations.

\subsection{Physics-informed loss function}

The weight vector~$\boldsymbol{\theta}$ defining the neural network~$\mathcal N^{\boldsymbol{\theta}}$ is found upon minimizing a suitable loss function. For a physics-informed neural network, such a loss function encapsulates the underlying initial--boundary value problem.

Consider an initial--boundary value problem for a general system of $L$ partial differential equations of order $n$ defined over the spatio-temporal domain $[0,t_{\rm f}]\times\Omega$, $\Omega\subset\mathbb{R}^d$, of the form
\begin{align}\label{eq:GeneralDifferentialEquation}
\begin{split}
&\Delta^l(t,\mathbf{x},\mathbf{u}_{(n)})=0,\quad l=1,\dots, L,\qquad t\in[0,t_{\rm f}],\ \mathbf{x}\in\Omega,\\[1ex]
&\mathsf{I}^{l_{\rm i}}(\mathbf{x},\mathbf{u}_{(n_{\rm i})}|_{t=0})=0,\quad l_{\rm i}=1,\dots, L_{\rm i},\qquad \mathbf{x}\in\Omega,\\[1ex]
&\mathsf{B}^{l_{\rm b}}(t,\mathbf{x},\mathbf{u}_{(n_{\rm b})})=0,\quad l_{\rm b}=1,\dots, L_{\rm b},\qquad t\in[0,t_{\rm f}],\ \mathbf{x}\in\partial\Omega,
\end{split}
\end{align}
where $t$ is the time variable, $\mathbf{x}=(x_1,\dots,x_d)$ is the tuple of spatial independent variables, $\mathbf{u}=(u^1,\dots, u^q)$ denotes the tuple of dependent variables, and $\mathbf{u}_{(n)}$ denotes the derivatives of the dependent variables with respect to the independent variables of order not greater than $n$. Here, $\mathsf{I}=(\mathsf{I}^1,\dots,\mathsf{I}^{L_{\rm i}})$ is the initial value operator and $\mathsf{B}=(\mathsf{B}^1,\dots,\mathsf{B}^{L_{\rm b}})$ denotes the boundary value operator. Thus, the initial value operator for a system of evolution equations is 
\[\mathsf{I} = \mathbf{u}|_{t=0}-\mathbf{f}(\mathbf{x}),\] 
with a fixed vector function $\mathbf{f}(\mathbf{x})=\big(f^1(\mathbf{x}),\dots, f^q(\mathbf{x})\big)$. In Dirichlet boundary conditions, 
\[\mathsf{B} = \mathbf{u}-\mathbf{g}(t,\mathbf{x})\] 
for a fixed vector function $\mathbf{g}(t,\mathbf{x})=\big(g^1(t,\mathbf{x}),\dots, g^q(t,\mathbf{x})\big)$.

Physics-informed neural networks aim to approximate the solution of system~\eqref{eq:GeneralDifferentialEquation} with a deep neural network. Denote by $\mathbf{u}^{\boldsymbol{\theta}}=\mathcal N^{\boldsymbol{\theta}}(t,\mathbf{x})$ the associated output of the neural network for the input values $t$ and $\mathbf{x}$. The weights~$\boldsymbol{\theta}$ are learned by minimizing the loss function
\begin{equation}\label{eq:compositeLossFunction}
\mathcal L(\boldsymbol{\theta}) = \mathcal L_\Delta(\boldsymbol{\theta}) + \gamma_{\rm i}\mathcal L_{\rm i}(\boldsymbol{\theta}) + \gamma_{\rm b}\mathcal L_{\rm b}(\boldsymbol{\theta}),
\end{equation}
over the neural network parameters~$\boldsymbol{\theta}$, where
\begin{align*}
\mathcal L_\Delta(\boldsymbol{\theta}) &= \frac{1}{N_\Delta}\sum_{i=1}^{N_\Delta}\sum_{l=1}^L\big|\Delta^l\big(t^i_\Delta,\mathbf{x}^i_\Delta,\mathbf{u}^{\boldsymbol{\theta}}_{(n)}(t^i_\Delta,\mathbf{x}^i_\Delta)\big)\big|^2,\\[1.5ex]
\mathcal L_{\rm i}(\boldsymbol{\theta}) &= \frac{1}{N_{\rm i}}\sum_{i=1}^{N_{\rm i}}\sum_{l_{\rm i}=1}^{L_{\rm i}}\big|\mathsf{I}^{l_{\rm i}}\big(\mathbf{x}^i_{\rm i},\mathbf{u}_{(n_{\rm i})}^{\boldsymbol{\theta}}(0,\mathbf{x}^i_{\rm i})\big)\big|^2,\\[1.5ex]
\mathcal L_{\rm b}(\boldsymbol{\theta}) &= \frac{1}{N_{\rm b}}\sum_{i=1}^{N_{\rm b}}\sum_{l_{\rm b}=1}^{L_{\rm b}}\big|\mathsf{B}^{l_{\rm i}}\big(t^i_{\rm b},\mathbf{x}^i_{\rm b},\mathbf{u}_{(n_{\rm b})}^{\boldsymbol{\theta}}(t^i_{\rm b},\mathbf{x}^i_{\rm b})\big)\big|^2
\end{align*}
are the PDE loss, the initial value loss and the boundary value loss, respectively, all of which are standard mean-squared errors, and $\gamma_{\rm i},\gamma_{\rm b}\in\mathbb{R}_{>0}$ are loss-weight parameters. Note that these losses are computed based on the collocation points $\big\{(t_\Delta^i,\mathbf{x}_\Delta^i)\big\}_{i=1}^{N_\Delta}$ for the system $\Delta$, the collocation points $\big\{(0,\mathbf{x}_{\rm i}^i)\big\}_{i=1}^{N_{\rm i}}$ for the initial data, and the collocation points $\big\{(t_{\rm b}^i,\mathbf{x}_{\rm b}^i)\big\}_{i=1}^{N_{\rm b}}$ for the boundary data, respectively.

By simultaneously minimizing all three loss contributions the system~\eqref{eq:GeneralDifferentialEquation} will be enforced along with the associate initial and boundary conditions, which will make sure that $\mathbf{u}^{\boldsymbol\theta}(t,\mathbf{x})$ will indeed be approximating the true solution of this system of differential equations.

What makes physics-informed neural networks particularly attractive is that the derivatives $\mathbf{u}^{\boldsymbol{\theta}}_{(n)}$ of neural network function approximator can be obtained using automatic differentiation~\cite{bayd18a}, which will not incur any discretization error as is typically the case in standard numerical approximations. Automatic differentiation produces exact derivatives with the only error incurring being due to finite-precision arithmetic.

Moreover, as automatic differentiation is used in deep learning toolkits such as \texttt{TensorFlow} for computing the derivatives of loss functions using the famous backpropagation algorithm~\cite{lecu15a}, automatic differentiation is readily available for computing the derivatives $\mathbf{u}^{\boldsymbol{\theta}}_{(n)}$, thereby alleviating the need for discretizing~\eqref{eq:GeneralDifferentialEquation} using traditional methods such as finite differences, finite volumes, finite elements or spectral methods.

While the above properties of physics-informed neural networks are indeed enticing, there is yet lacking a formal theory regarding their convergence for arbitrary (nonlinear) systems of differential equations. Experience shows that physics-informed neural networks do yield meaningful and accurate results for systems of differential equations arising in a wide variety of fields, see e.g.~\cite{jin21a,kiss20a,mao20a,rais19a}, but whether or not they are guaranteed to converge to the true solution of the underlying system of differential equations is not yet known; indeed, they may even fail to be trained successfully.

Recently, the paper~\cite{shin20a} shed some light on the estimation error of physics-informed neural networks. In general, the error of a physics-informed neural network is composed of the \textit{approximation error} (the error of the network in approximating the true solution, which is well understood through the Universal Approximation Theorem), the \textit{optimization error} (the error of the optimization algorithm in finding the global minimum of the loss function), and the \textit{estimation error} (the error in representing the chosen function class due to using only finitely many collocation points). While the issue of finding the global minima in neural networks is highly nontrivial due to the loss function being nonconvex, the paper \cite{shin20a} indeed showed consistency of physics-informed neural networks by proving that as the number of data points grows, the estimation error goes to zero.

Similarly, in~\cite{wang20a} the question of when physics-informed neural networks fail to learn successfully was investigated. Specifically, the paper~\cite{wang20a} investigated the so-called ``spectral bias'' of physics-informed neural networks, which concerns their inability to learn functions with high frequencies. The spectral bias was discussed in~\cite{raha19a}, where it was shown that lower frequency functions are learned first in deep neural networks, and that these frequencies are more stable under random perturbations of the network parameters than high-frequency functions. To improve the training of physics-informed neural networks, the paper~\cite{wang20a} proposed a dynamic weighting strategy for the various components of the physics-informed loss function, i.e.\ by adaptively changing $\gamma_{\rm i}$ and $\gamma_{\rm b}$ over the training period, see also~\cite{wang20b} for a further discussion on this issue.

We note that for the present study the accurate representation of high-frequency function is neither necessary nor desirable. Indeed, dynamic meteorology is plagued by the presence of high frequency waves, which typically have to be filtered out before a numerical implementation of the governing equations of hydro-thermodynamics, see e.g.~\cite{sato04a}. Still, there are some adaptations of the basic physics-informed neural network training algorithm that we found necessary for the shallow-water equations, which are detailed in the subsequent section.

\section{The shallow-water equations on the sphere}\label{sec:PINNsSWE}

Here we present the form of the shallow-water equations as used for the implementation in physics-informed neural networks.

\subsection{Non-dimensionalization and scaling}

The shallow-water equations on the sphere in latitude--longitude coordinates are
\begin{align}\label{eq:ShallowWaterSphere}
\begin{split}
&u_t + \frac{u}{a\cos\theta}u_\lambda + \frac{v}{a}u_\theta -\left(f+\frac{u}{a}\tan\theta\right)v + \frac{g}{a\cos\theta} h_\lambda = 0, \\[.8ex]
&v_t + \frac{u}{a\cos\theta}v_\lambda + \frac{v}{a}v_\theta +\left(f+\frac{u}{a}\tan\theta\right)u + \frac{g}{a} h_\theta = 0, \\[.3ex]
&h^*_t + \frac{u}{a\cos\theta}h^*_\lambda + \frac{v}{a}h^*_\theta + \frac{h^*}{a\cos\theta} (u_\lambda + (v\cos\theta)_\theta) = 0,
\end{split}
\end{align}
where $u$ and $v$ are the $\lambda$- and the $\theta$-components of the velocity vector in the spherical coordinates, respectively, $h=h^*+b$ is the total water height, with $b$ being the height of the bottom topography. The radius of the Earth is denoted by $a$, $f=2\omega\sin\theta$ is the Coriolis parameter with $\omega$ being the angular velocity of the Earth. See e.g.~\cite{sato04a,will92Ay} for a discussion of these equations along with a presentation of different coordinate systems being used besides the standard latitude--longitude coordinates.

We consider initial boundary-value problems for the shallow-water equations on the entire sphere, $(t,\lambda,\theta)\in\Omega=[0,t_{\rm f}]\times[-\pi,\pi]\times[-\pi/2,\pi/2]$, which implies the periodic boundary conditions with respect to the longitude and the obvious continuity conditions on the poles. Given the vastly different ranges of the variables $u$ and $v$, ranging to values of the order of $10^2$, and $h$, ranging to values of the order of $10^4$, we find it essential to non-dimensionalize the system~\eqref{eq:ShallowWaterSphere} before training physics-informed neural networks, see~\cite{glor10a,kiss20a} for further discussions on this issue.

We non-dimensionalize the shallow-water equations by introducing the following characteristic scales of large-scale atmospheric phenomena,
\[
L = a\ [{\rm m}],\quad T=1\ [{\rm day}]=86400\ [{\rm s}],\quad U = L/T\ [{\rm m{\cdot}s}^{-1}],\quad H=c\frac{L^2}{T^2g}\ [{\rm m}],
\]
with $c$ being an arbitrary non-dimensional constant. Introducing the re-scaled variables $t=T\hat t$, $\lambda=\hat \lambda$, $\theta=\hat\theta$, $u=U\hat u$, $v=U\hat v$, $h^*=H\hat h$ and $b^*=H\hat b$, we can rewrite system~\eqref{eq:ShallowWaterSphere} as the non-dimensional system
\begin{align}\label{eq:ShallowWaterSphereNonDimensional}
\begin{split}
&\hat u_{\hat t} + \frac{\hat u}{\cos\hat\theta}\hat u_{\hat\lambda} + \hat v\hat u_{\hat\theta} -\left(\hat f+\hat u\tan\hat\theta\right)\hat v + \frac{c}{\cos\hat\theta}\hat h_{\hat\lambda} = 0, \\[0.5ex]
&\hat v_{\hat t} + \frac{\hat u}{\cos\hat\theta}\hat v_{\hat\lambda} + \hat v\hat v_{\hat\theta} +\left(\hat f+\hat u\tan\hat\theta\right)\hat u + c \hat h_{\hat\theta} = 0, \\
&\hat h^*_{\hat t} + \frac{\hat u}{\cos\hat \theta}\hat h^*_{\hat\lambda} + \hat v\hat h^*_{\hat\theta} + \frac{\hat h^*}{\cos\hat\theta} (\hat u_{\hat\lambda} + (\hat v\cos\hat\theta)_{\hat\theta}) = 0,
\end{split}
\end{align}
where $\hat f=Tf$ is the non-dimensional Coriolis parameter. In the following, we set $c=1$ for the global steady state test case, and $c=20$ for the flow over the mountain and the Rossby--Haurwitz wave test case. This was found experimentally and ensures that all variables $(\hat u, \hat v, \hat h)$ are of unit magnitude. Below we omit the hats over variables and refer to system~\eqref{eq:ShallowWaterSphereNonDimensional} by $\Delta=0$.

\subsection{Neural network architectures and collocation points}

A crucial aspect in solving differential equations with neural networks is the right choice for the network architecture, i.e., choosing the proper number of hidden layers, units per hidden layer and activation functions. As was indicated in Section~\ref{sec:TheoryPINNsSWE}, there is yet no unifying theory regarding the convergence of physics-informed neural networks, that could guide in the design of the appropriate neural networks.

However, in the previous works~\cite{jin21a,kiss20a,mao20a,rais19a} a variety of physics-informed networks have been trained successfully, which employ neural network architectures ranging from 2 to 12 hidden layers, with 20 to 200 units per hidden layer. One potential approach in selecting a specific architecture is akin to the standard approach in machine learning/deep learning, which concerns the bias--variance trade-off, see e.g.~\cite{gero19a}: If the model is too simple to accurately capture the training data, then the network may not have enough capacity, in which case the number of hidden layers and/or units per layer has to be increased. If the model has a low training error but fails to accurately model the solution throughout the computational mesh then the network may have too much capacity or the number of supplied collocation points is not enough. Then either reducing the capacity of the network again by reducing the number of hidden layers and/or units per layer or increasing the number of collocation points may alleviate this overfitting issue, thus yielding better generalization abilities of the network. Since training physics-informed neural networks in general is not overly time-consuming, a standard hyperparameter grid search typically yields a suitable architecture in a limited amount of computational time.

This is why in the test cases below we adapt the architecture and/or number of collocation points for the computational domain. For example, the linear advection test case (Section~\ref{sec:AdvectionCosineBell}) does not require as deep a neural network as the Rossby--Haurwitz wave test case (Section~\ref{sec:RossbyHaurwitzWave}). For further guidance on the relation between the neural network width/depth and the accuracy of the numerical results we include a small convergence study in this regard on the nonlinear zonal geostrophic flow test case (Section~\ref{sec:NonlinearGeostrophicFlow}).

As activation functions we choose the hyperbolic tangent function as it is differentiable over its entire domain of definition, in contrast to most of the more popular ReLU variants being used in deep learning. We experimentally found the hyperbolic tangent activation function to give good results. For a more in-depth study on the problem of activation functions in PINNs, see~\cite{jagt20b}.

The collocation points $\big\{(t_\Delta^i,\lambda_\Delta^i,\theta_\Delta^i)\big\}_{i=1}^{N_\Delta}$ for enforcing the system $\Delta=0$,
which samples the spatio-temporal domain $\Omega=[0,t_{\rm f}]\times[-\pi,\pi]\times[-\pi/2,\pi/2]$, are chosen in such a way
that their counterparts in $[0,t_{\rm f}]\times S^2$ are uniformly distributed, with fewer points closer to the poles, and more points closer to the equator.
Here $S^2$ denotes the sphere.
More specifically, we first sample an $(N_\Delta,3)$-array of points from the interval $[0,1]$ using Latin hypercube sampling~\cite{stei87a}.
Then each point, $(t^*,\lambda^*,\theta^*)$, from the obtained set is mapped to~$\Omega$ by the point transformation
\[
t_\Delta = t_{\rm f} t^*,\quad \lambda_\Delta = -\pi + 2\pi\lambda^*,\quad \theta_\Delta = -\frac{\pi}{2} + \arccos(1-2\theta^*),
\]
cf.~\cite{simo15a}.
The chosen transformation for~$\theta$ ensures the uniform distribution of the respective points in $[0,t_{\rm f}]\times S^2$.
For a visual representation of the collocation points being used, see Fig.~\ref{fig:MultiModel} (left).

We explicitly enforce the boundary conditions on the sphere as hard constraint. This is done by feeding the inputs in the original $(\lambda,\theta)$-coordinates into a coordinate-transform layer first that outputs $\mathbf{r}=\mathbf{r}(\lambda,\theta)=(\cos\theta\cos\lambda,\cos\theta\sin\lambda,\sin\theta)$, i.e., that associates to each $(\lambda,\theta)$-pair the Cartesian coordinates $\mathbf{r}=(x,y,z)$ via embedding the sphere into $\mathbb{R}^3$. In other words, the constructed neural networks learn the mappings
$
\mathbf{u}^{\boldsymbol{\theta}}=\mathbf{u}^{\boldsymbol{\theta}}(t,\mathbf{r}(\lambda,\theta)),\ (t,\lambda,\theta)\in\Omega,
$
where $\mathbf{u}^{\boldsymbol{\theta}}=(u^{\boldsymbol{\theta}},v^{\boldsymbol{\theta}},h^{\boldsymbol{\theta}})$,
and $\boldsymbol{\theta}$ is the weight vector.
This way, the loss function to be minimized has contributions only from the PDE loss~$\mathcal L_{\Delta}$ and the initial value loss~$\mathcal L_{\rm i}$.

\subsection{Optimization issues}

Composite loss functions such as those being used for physics-informed neural networks are challenging since multiple tasks have to be solved at the same time. In the language of machine learning, physics-informed neural networks are what is called a \textit{multi-task learning problem}~\cite{sene18a,yu20a}. In multi-task learning it is well known that simply adding the loss functions associated to each task may not lead to each task being learned properly, which may explain some of the practical training difficulties observed for physics-informed neural networks.

Instead of heuristically tuning the coefficients in the weighted summation loss~\eqref{eq:compositeLossFunction} as being used in physics-informed neural networks, with each loss component potentially corresponding to a conflicting objective for the optimization algorithm, a solution for the multi-task learning problem could alternatively be found by achieving \textit{Pareto optimality}~\cite{sene18a}. In particular, it was shown in~\cite{sene18a} that for the special case of two objective functions, the solution to the optimization problem
\[
\min_{\alpha\in[0,1]}\big\|\alpha\nabla_{\boldsymbol{\theta}}\mathcal L_\Delta + (1-\alpha)\nabla_{\boldsymbol{\theta}}\mathcal L_{\rm i}\big\|_2^2
\]
is given by
\[
\hat \alpha = \left[\frac{(\nabla_{\boldsymbol{\theta}}\mathcal L_{\rm i}-\nabla_{\boldsymbol{\theta}}\mathcal L_\Delta)^{\rm T}\nabla_{\boldsymbol{\theta}}\mathcal L_{\rm i}}{\big\|\nabla_{\boldsymbol{\theta}}\mathcal L_\Delta-\nabla_{\boldsymbol{\theta}}\mathcal L_{\rm i}\big\|_2^2}\right]_{\rm c},
\]
where $[\cdot]_{\rm c}$ represents clipping to $[0,1]$, $[\cdot]_{\rm c}=\max(\min(\cdot,1),0)$. Note that since it is always possible to enforce the boundary conditions in a physics-informed neural network as a hard constraint, the above solution for two objective functions given by the PDE loss and the initial value loss should be sufficient for most practical cases.\footnote{For more than two objectives, the paper~\cite{sene18a} presents an iterative solution algorithm.} This solution also provides a theoretically justified alternative to the various heuristic loss scaling strategies for physics-informed neural networks that have been presented in the past.

As a further alternative for a solution of the multi-task learning problem presented in~\cite{sene18a}, the paper~\cite{yu20a} introduces a more heuristic strategy. In particular, they identified the \textit{tragic triad} consisting of \textit{conflicting gradients}, \textit{dominating gradients} and \textit{high curvature} as the culprit for optimization issues in gradient descent methods for multi-task learning. This is in line with the work~\cite{wang20b} that focused on the issue of dominating gradients. The solution to the tragic triad problem put forth in~\cite{yu20a} is a simple \textit{projecting conflicting gradients} (PCGrad) strategy. That is, if the gradients of the loss function components are found to point in opposing directions then these gradients are altered by projecting each onto the normal plane of the other, which will prevent the gradient descent algorithm from stalling in suboptimal minima.

We have implemented both strategies discussed above and found that the first strategy often leads to excessively small (or large) values of $\alpha$, which implies that the optimizer will make most progress by nearly ignoring one part of the loss function, which means that a large number of gradient descent steps have to be taken until an accurate solution is found. In turn, the second method seems to overcome suboptimal local minima much more quickly. This is why we only report the results for the second method below.

\subsection{Longer training times}

One of the drawbacks of physics-informed neural networks is that the number of collocation points required continuously increases the larger the domain of the problem is. As a solution to this problem, in~\cite{meng20a} an adaptation of parallel-in-time algorithms for physics-informed neural networks was proposed. Here, in an iterative process a fast coarse-grained PINN solver over a larger time interval is used to guide many independent, parallel, and accurate fine-scale PINN solvers to improve the coarse-grained solution.

While the perspective of using a parallel-in-time algorithm for longer integrations of partial differential equations with physics-informed neural networks is appealing, it relies on the availability of a suitable coarse-grained PINN solver, which has to be designed to solve a simplified problem instead of the full system of partial differential equations at hand. In~\cite{meng20a} this was illustrated for problems that can clearly be separated in an ``easy'' and a ``hard'' part, which made choosing the simplified model for the coarse-grained PINN solver straightforward. This may not be as straightforward for the shallow-water equations.

\begin{figure}[!ht]
\centering
\begin{subfigure}{0.43\textwidth}
  \centering
  \includegraphics[width=\linewidth]{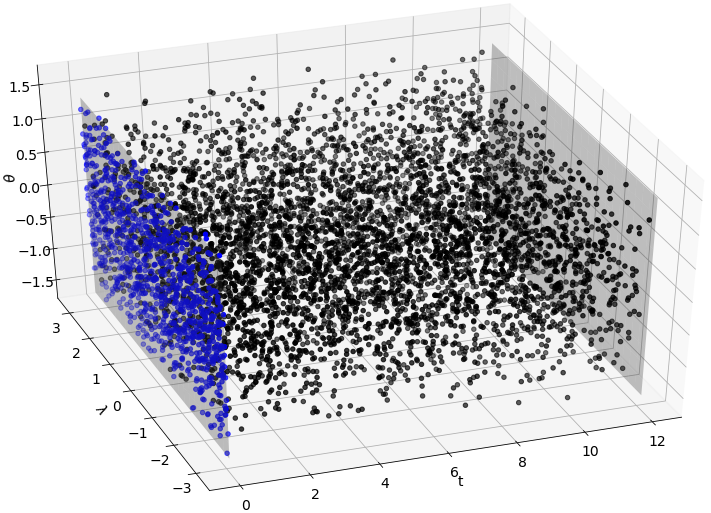}
\end{subfigure}
\hfil
\begin{subfigure}{0.43\textwidth}
  \centering
  \includegraphics[width=\linewidth]{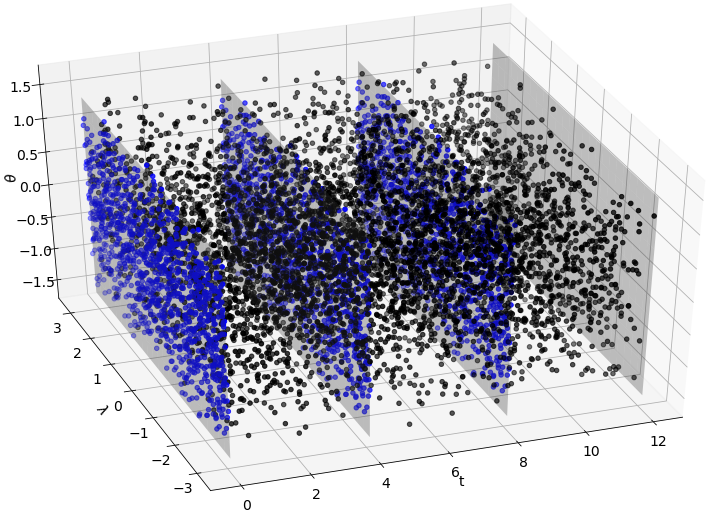}
\end{subfigure}
\caption{Pictorial representation of the multi-model approach for physics-informed neural networks. \textit{Left:} Collocation points for a standard single-model physics-informed neural network. \textit{Right:} Multi-model approach. The total integration interval $[0,12]$ is divided into 3 non-overlapping sub-intervals of equal length and one new physics-informed neural network is trained for each sub-interval sequentially. The initial conditions for the $(i+1)$st model are obtained by evaluating the $i$th model at the right boundary of its time interval. Black points are collocation points for computing $\mathcal L_\Delta$, blue points are initial value points for computing $\mathcal L_{\rm i}$.}\label{fig:MultiModel}
\end{figure}

As an alternative, we propose the following strategy here: We split the time interval $[0,t_{\rm f}]$ into $n$ non-overlapping subintervals $[0,t_1]$, $[t_1,t_2]$, \dots, $[t_{n-1},t_{\rm f}]$ and solve the shallow-water equations on each subinterval consecutively, by training a new neural network for each subinterval. The solution at the right boundary of the time interval as computed by the $i$th neural network then becomes the initial condition for $(i+1)$st neural network, see Figure~\ref{fig:MultiModel} for a visual representation.  We also initialize the weights of the $(i+1)$st network using the trained weights of the $i$th network, and found experimentally that this allows us to train each subsequent network using substantially fewer epochs than are normally required to train physics-informed neural networks

There are a few advantages of this approach. Firstly, it is possible to continue training a physics-informed neural network that has been trained for the interval $[0,t_{\rm f}]$ for the longer interval $[0,t_{\rm f}']$, where $t_{\rm f}'>t_{\rm f}$, without having to start the training procedure from scratch. Secondly, it would be possible to use neural networks of varying size for different time subintervals, and thereby tailor these networks to the underlying evolution of the solution of the given system of differential equations. As an example, the transition of laminar to turbulent flow could thus naturally be split into several networks, one that learns to simulate laminar flow, one that learns the transition, and one or more for the turbulent phase of the evolution.

This is in line with modern deep learning that recognizes that it is better to split a complicated task into many subtasks and to train neural networks for each subtask individually instead of training one network for the entire task (such as in neural language translation, where encoder--decoder architectures are used, with the encoder learning to understand text in one language and the decoder learning to translate the outputs of the encoder into another language)~\cite{gero19a}.

A related idea has been developed in~\cite{jagt20c}, where an extended parallel space--time domain decomposition (XPINN) was proposed to improve the performance of vanilla physics-informed neural networks. The main difference of XPINN to our proposed multi-model approach is that we train our sub-models sequentially and independently, and therefore do not have to take into account additional interface and boundary conditions between neighboring subdomains. This in particular simplifies the form of the loss function to be minimized which in turn reduces the computational cost of the optimization procedure, in particular as we include the boundary condition as a hard constraint directly in the neural network architecture. Owing to the parallel nature of XPINN, weight initialization of subsequent models using previously trained models is therefore also not possible. In contrast, we do expect XPINN to be advantageous over the proposed multi-model approach for large-scale spatio-temporal problems, which are not considered in the present paper, or if multiple GPUs are available for training. For a more in-depth discussion on the theoretical error bounds of XPINN, and a discussion on when this method is guaranteed to outperform standard physics-informed neural networks, consult reference~\cite{hu21a}.

We should also like to note that we were not able to train the physics-informed neural networks presented below with any of the training strategies proposed in~\cite{sene18a,wang20b,wang20a,yu20a} for the entire temporal domain $[0,t_{\rm f}]$ at once unless an excessive amount of collocation points was used. Scaling up the number of collocation points even further may be infeasible for more complicated problem, as it will increase the training time of these networks substantially and/or place a more substantial burden on the available GPU resources.

\subsection{Implementation}

The neural networks have been implemented and trained using \texttt{TensorFlow} 2.4\footnote{The code is available from \url{https://github.com/abihlo/pinnsSWE}.}. Training was done on a single NVIDIA V100 GPU, with each model taking roughly 5 minutes to 5 hours to train when using the lowest to highest number of collocation points being reported below. The loss function is minimized using the popular \textit{Adam optimizer}~\cite{king14a} for training all networks. We have further experimented with various variants of the Adam optimizer, such as Nesterov-accelerated Adam~\cite{doza16a}, Lookahead~\cite{zhan19b} and Lamb~\cite{you19a}. Some of these optimizers were able to find a better solution than Adam for some of the benchmarks below, but none of them found a better solution for all benchmarks. Therefore, we report results of Adam alone, as this is still the most prominent optimizer to date, and it is the one traditionally being used for training physics-informed neural networks. Note that Adam is a first-order optimizer and we do not use second-order L-BFGS method for fine-tuning the Adam pre-trained neural networks as done, e.g., in~\cite{rais19a}, since as of early 2021 \texttt{TensorFlow} 2.x did not provide a canonical interface to L-BFGS as implemented, e.g.\ in \texttt{scipy.optimize}.

\looseness=-1
While all computations have been carried out in standard (double) precision floating point arithmetic, we should like to indicate the potential avenue of reducing the numerical precision without significantly increasing the overall errors. Some initial testing with our code has shown that the quality of the results reported does not significantly degrade when reduced (single) precision arithmetic is being used. This result is in line with the results reported in~\cite{coop20a}, which have shown that linearly reducing the precision being used does not significantly affect the forecast skills. Pursuing this avenue may enable larger simulations at reduced computational cost, which may be essential as the proposed algorithm scales, leaving mini-batch sizes the same, as $N_{\Delta}^3$.

\section{Results}\label{sec:ResultsPINNsSWE}

A standard test suite for numerical approximations to the shallow-water equations on the sphere was proposed in~\cite{will92Ay}. Here we report the results of our neural networks for some of these test cases. The global constants for all test cases are
\[
a=6.37122\cdot 10^{6}\ [{\rm m}],\qquad \omega=\frac{2\pi}{86400}\ [{\rm s}^{-1}],\qquad g=9.80616\ [{\rm m{\cdot}s}^{-1}],
\]
corresponding to the radius of Earth, the angular velocity of Earth's rotation, and the gravitational acceleration.

Unless otherwise specified, we use mini-batch gradient descent for all experiments reported below with a mini-batch size of $1000$ collocation points for evaluating~$\mathcal L_\Delta$ and $100$ points for evaluating~$\mathcal L_{\rm i}$.
Note that the presented results are obtained using a variety of neural network architectures of different complexity. In particular, we have strived to use the simplest architecture that gives a reasonably accurate solution for each of the benchmarks proposed in~\cite{will92Ay}, which themselves are of different complexity.

\looseness=-1
From the four test cases reported below, only the first and the second have known exact solutions, which we employ for measuring errors in the respective test cases. For the last two test cases, we report errors as measured against reference numerical solutions that are obtained by applying the variational integrator developed for the shallow-water equations on the sphere in~\cite{brec19a}.

\subsection{Advection of a cosine bell around the sphere}\label{sec:AdvectionCosineBell}

The initial condition for this test case is
\begin{align}\label{eq:Williamson1}
\begin{split}
&u= u_0(\cos\theta\cos\alpha+\sin\theta\cos\lambda\sin\alpha),\\
&v= -u_0\sin\lambda\sin\alpha,\\
&h(\lambda,\theta) = \left\{\begin{array}{ccc}
h_0\big(1+\cos(\pi r/R)\big)/2 & \textup{if}& r<R,\\
0 & \textup{if}& r\geqslant R,
\end{array} \right.
\end{split}
\end{align}
where $r=a\arccos\big(\sin\theta_{\rm c}\sin\theta+\cos\theta_{\rm c}\cos\theta\cos(\lambda-\lambda_{\rm c})\big)$. The constants are chosen as $u_0=2\pi a/12\ {\rm m}{\cdot}{\rm days}^{-1}$, $h_0=1000\ {\rm m}$, $R=a/3$, and the bell center is $(\lambda_{\rm c},\theta_{\rm c})=(3\pi/2,0)$.

This test case does not solve the full shallow-water equations, but rather just tests the advection of the height equation in~\eqref{eq:ShallowWaterSphere}. In other words, the equations for $u$ and $v$ are ignored and the analytical wind field in~\eqref{eq:Williamson1} is being chosen, which due to being non-divergent turns~\eqref{eq:ShallowWaterSphere} into a linear advection equation for $h$,
\[
 h_t + \frac{u}{a\cos\theta}h_\lambda + \frac{v}{a}h_\theta=0.
\]
Note that since $b=0$ we have $h=h^*$. The solution shape for $h$ specified in~\eqref{eq:Williamson1} should thus be advected around the Earth in $t_{\rm f}=12$ days without changing its shape.

We test two orientations of the advecting wind, one time with $\alpha=0$, which corresponds to advection along the equator, and one time with $\alpha=\pi/2$, which corresponds to advection over the poles.

Therefore, the neural network architecture for this case accepts $(t,\lambda,\theta)$ as input and outputs the single variable $h_{\boldsymbol{\theta}}$. We choose a neural network consisting of 4 hidden layers with 20 units per layer and with $\tanh$ as activation function. This yields a total of 1380 weight and bias parameters that the network has to optimize. The learning rate of the Adam optimizer was set to $\eta=10^{-3}$ for all experiments.

\begin{figure}[!ht]
\centering
\begin{subfigure}{0.33\textwidth}
  \centering
  \includegraphics[width=\linewidth]{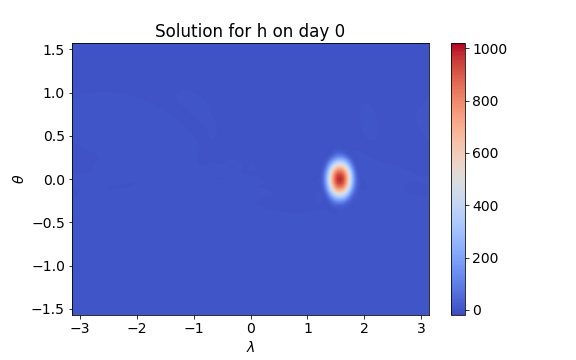}
\end{subfigure}
\begin{subfigure}{0.33\textwidth}
  \centering
  \includegraphics[width=\linewidth]{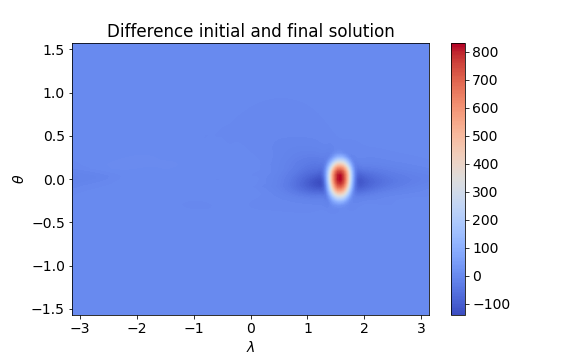}
\end{subfigure}
\begin{subfigure}{0.32\textwidth}
  \centering
  \includegraphics[width=\linewidth]{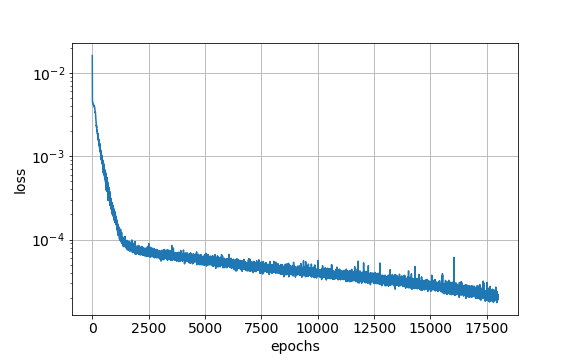}
\end{subfigure}
\begin{subfigure}{0.33\textwidth}
  \centering
  \includegraphics[width=\linewidth]{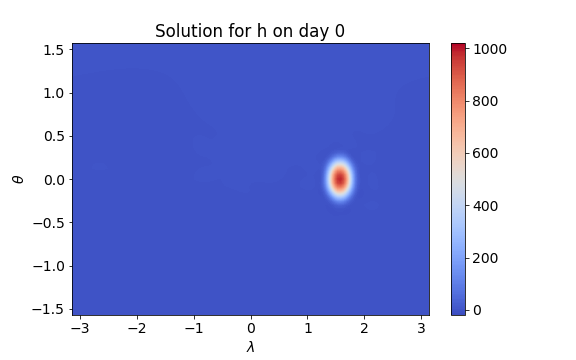}
\end{subfigure}
\begin{subfigure}{0.33\textwidth}
  \centering
  \includegraphics[width=\linewidth]{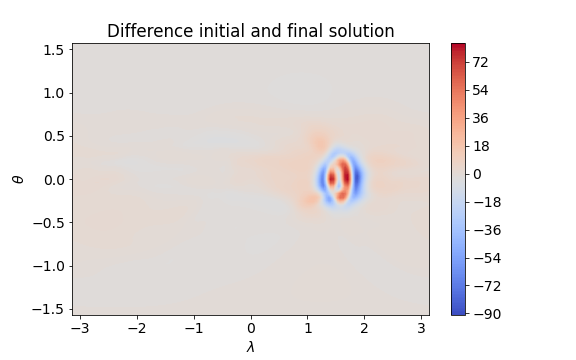}
\end{subfigure}
\begin{subfigure}{0.32\textwidth}
  \centering
  \includegraphics[width=\linewidth]{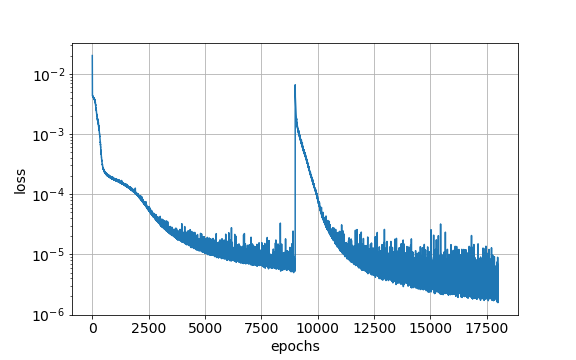}
\end{subfigure}
\begin{subfigure}{0.33\textwidth}
  \centering
  \includegraphics[width=\linewidth]{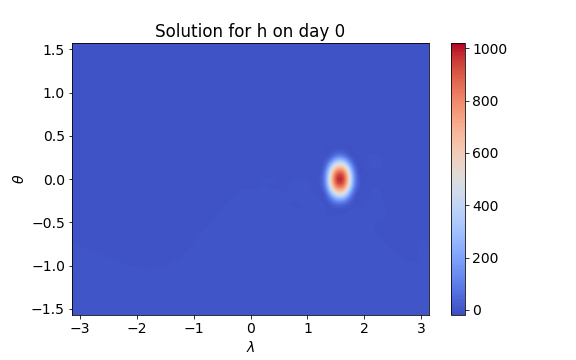}
\end{subfigure}
\begin{subfigure}{0.33\textwidth}
  \centering
  \includegraphics[width=\linewidth]{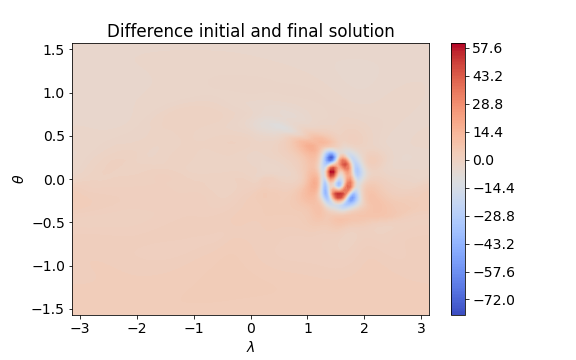}
\end{subfigure}
\begin{subfigure}{0.32\textwidth}
  \centering
  \includegraphics[width=\linewidth]{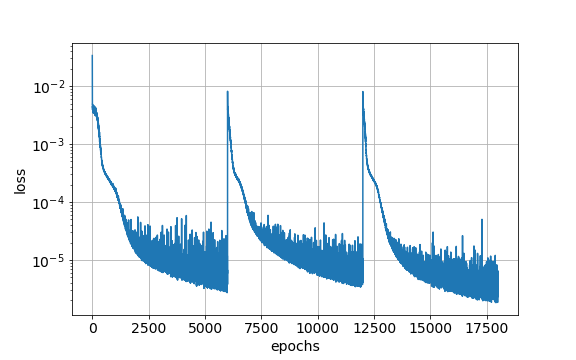}
\end{subfigure}
\begin{subfigure}{0.33\textwidth}
  \centering
  \includegraphics[width=\linewidth]{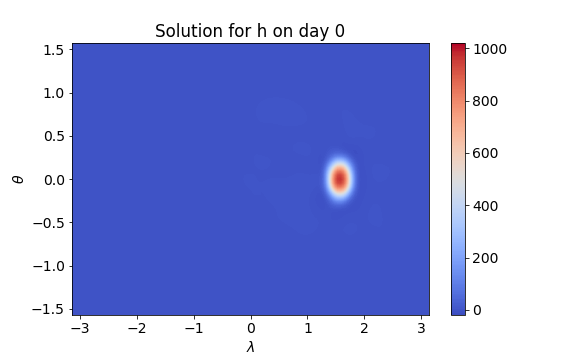}
\end{subfigure}
\begin{subfigure}{0.33\textwidth}
  \centering
  \includegraphics[width=\linewidth]{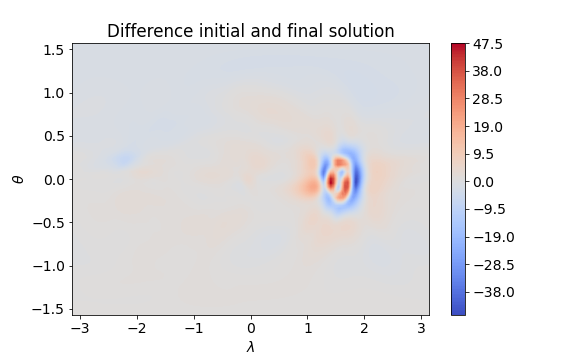}
\end{subfigure}
\begin{subfigure}{0.32\textwidth}
  \centering
  \includegraphics[width=\linewidth]{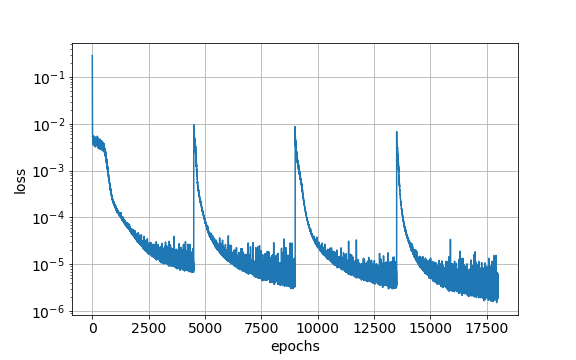}
\end{subfigure}
\caption{Results on the cosine bell advection test case proposed in~\cite{will92Ay} for $\alpha=0$. \textit{Top to bottom:} Solving this test case with 1, 2, 3 or 4 separate neural networks. The top row thus corresponds to the standard physics-informed neural network case. \textit{Left to right:} Reconstructions of the initial condition, differences between the initial condition and the solution on day 12, loss functions.}
\label{fig:WilliamsonTest1}
\end{figure}

We begin with the case of $\alpha=0$, which should yield a linear advection of the initial cosine bell along the equator. To test the multi-model approach we carry out four sets of experiments, using one, two, three or four models being trained for a total of $18000$ epochs per experiment. That is, when using only one model for the entire interval $[0,t_{\rm f}]$ this model is being trained for $18000$ epochs. When using two models on $[0,t_{\rm f}/2]$ and $[t_{\rm f}/2,t_{\rm f}]$, respectively, each model is being trained for $9000$ epochs, etc. The same total number of collocation points are being used for each experiment, which is $N_{\rm i}=10^4$ for the initial values, and $N_\Delta=10^5$ to cover the entire spatio-temporal domain $\Omega=[0,t_{\rm f}]\times[-\pi,\pi]\times[-\pi/2,\pi/2]$. Note that this means that the multi-model experiments get progressively faster to run, since each model is used at only a fraction of the total number of epochs, which each epoch consisting of only a fraction of the total number of points $N_\Delta$. The results of these experiments are depicted in Figure~\ref{fig:WilliamsonTest1}.

Note that while in all experiments the physics-informed networks have learned to accurately capture the initial condition, as expected the single model results are by far the worst among all experiments, as it exhibits an excessive amount of numerical diffusion due to the inability of the physics-informed neural network to properly link the initial condition to the temporal evolution of the initial cosine bell over the entire interval $[0,t_{\rm f}]$. Indeed, the solution is essentially dissipated completely here. Using at least two models gives a sufficiently accurate solution, with more than two models yielding even better levels of overall error, which in addition are obtained at a fraction of the computational cost of the experiments with fewer models. The time series of the loss functions also indicate that while starting to train a new model substantially increases the loss function (due to a miss-match of the old and new model's initial conditions) for short periods of time, the overall loss still decreases rapidly again after a few thousand epochs, with more rapid decrease the more models are being used. This is due to each model only being required to fit a decreasing number of collocation points the more models are being used, which considerably speeds up learning.

\begin{figure}[!ht]
${ }$\\[2ex]
\centering
\begin{subfigure}{0.33\textwidth}
  \centering
  \includegraphics[width=\linewidth]{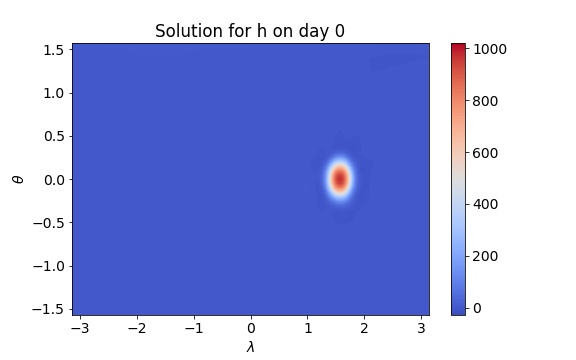}
\end{subfigure}
\begin{subfigure}{0.33\textwidth}
  \centering
  \includegraphics[width=\linewidth]{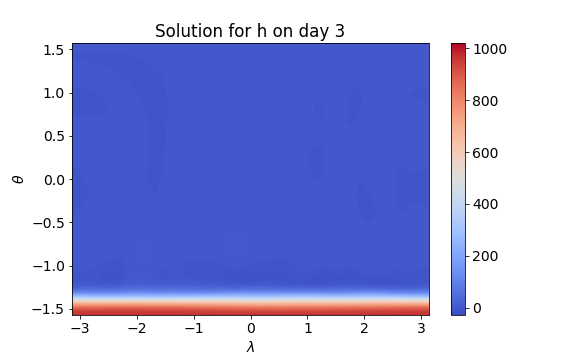}
\end{subfigure}
\begin{subfigure}{0.32\textwidth}
  \centering
  \includegraphics[width=\linewidth]{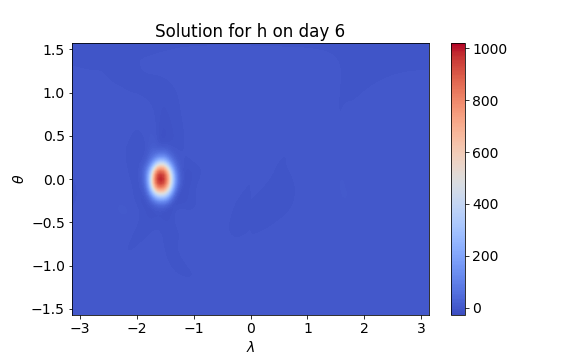}
\end{subfigure}
\\[3ex]
\begin{subfigure}{0.33\textwidth}
  \centering
  \includegraphics[width=\linewidth]{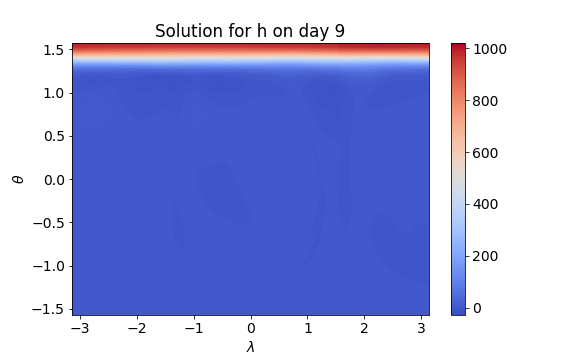}
\end{subfigure}
\begin{subfigure}{0.33\textwidth}
  \centering
  \includegraphics[width=\linewidth]{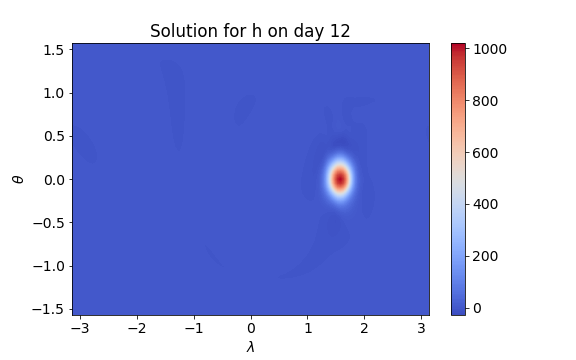}
\end{subfigure}
\begin{subfigure}{0.32\textwidth}
  \centering
  \includegraphics[width=\linewidth]{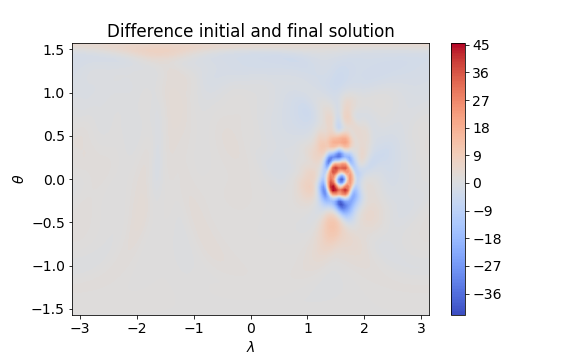}
\end{subfigure}
\\[2ex]${ }$
\caption{Results on the cosine bell advection test case proposed in~\cite{will92Ay} for $\alpha=\pi/2$. \textit{Top to bottom:} Solution at days $0$, $3$, $6$, $9$ and $12$ as well as difference between the solution at days $0$ and $12$.}\vspace{2ex}
\label{fig:WilliamsonTest1Pole}
\end{figure}

We next consider the case of $\alpha=\pi/2$, which corresponds to advection of the initial cosine bell over the equator. In latitude--longitude coordinates this is a challenging test case due to convergence of the meridional lines at the poles. We use the same setup as for the case of $\alpha=0$ using four models to obtain the solution over the time interval $[0,t_{\rm f}]$. The results of this experiment are depicted in Figure~\ref{fig:WilliamsonTest1Pole}. As is evidenced by these numerical results, despite the singularity in the latitute--longitude coordinate system, the physics-informed neural network is capable of advecting the cosine bell over both poles with comparable accuracy as in the previous case where we considered advection along the equator.

In an attempt to speed up the convergence of the training of the physics-informed neural networks we implemented the method of adaptive activation functions proposed in~\cite{jagt20b}. However, the results we obtained upon replacing the $\tanh$-activation function with an adaptive version yielded quite similar results. Thus, in what follows we only use fixed activation functions.

\subsection{Global steady state nonlinear zonal geostrophic flow}\label{sec:NonlinearGeostrophicFlow}

\begin{figure}[!ht]
\centering
\begin{subfigure}{0.33\textwidth}
  \centering
  \includegraphics[width=\linewidth]{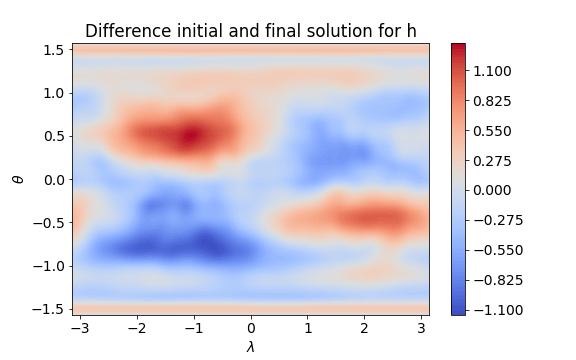}
\end{subfigure}
\begin{subfigure}{0.33\textwidth}
  \centering
  \includegraphics[width=\linewidth]{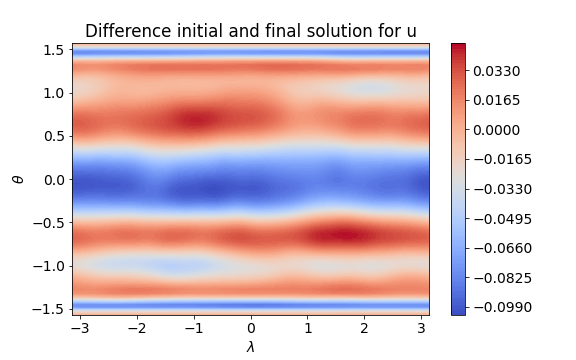}
\end{subfigure}
\begin{subfigure}{0.32\textwidth}
  \centering
  \includegraphics[width=\linewidth]{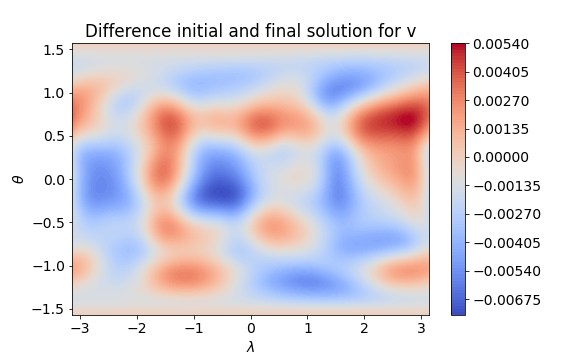}
\end{subfigure}
\caption{Results on the nonlinear zonal geostrophic flow test case~\cite{will92Ay} for $\alpha=0$. \textit{Left to right:} Difference between solution at days 0 and 5 for $h$, $u$ and $v$, whereas $h\in[1000,3000]$, $|\mathbf{v}|\in[0,40]$.}
\label{fig:WilliamsonTest2}
\end{figure}

This is the first test case that considers the full nonlinear shallow-water equations, and the only test case for which an exact solution is known. The initial condition for this test case is
\begin{align}\label{eq:Williamson2}
\begin{split}
&u= u_0(\cos\theta\cos\alpha+\sin\theta\cos\lambda\sin\alpha),\\[0.5ex]
&v= -u_0\sin\lambda\sin\alpha,\\
&gh = gh_0 -\left(a\omega u_0+\frac{u_0^2}{2}\right)\sin^2\theta,
\end{split}
\end{align}
with the parameter values $u_0=2\pi a/12\ {\rm m}{\cdot}{\rm days}^{-1}$ and $gh_0=2.94\cdot 10^4\ {\rm m}^2{\cdot}{\rm s}^{-2}$. This is a steady-state solution of the shallow-water equations~\eqref{eq:ShallowWaterSphere}. The Coriolis parameter for this solution is
\[
f = 2\omega(\sin\theta\cos\alpha-\cos\lambda\cos\theta\sin\alpha),
\]
and this test is to be run for 5 days. We again consider the case $\alpha=0$. The physics-informed neural network solutions using neural networks with 8 layers and 80 units per layer is presented in Figure~\ref{fig:WilliamsonTest2}. We used $N_\Delta=10^5$ collocation points and $N_{\rm i}=10^4$ initial points.

To better understand the influence of the neural network architectures on the approximation errors obtainable we carry out a small hyperparameter (convergence) study varying the number of hidden layers and number of units per layers, along with the number of initial and collocation points. This convergence study is also carried out for $\alpha=0$. We follow~\cite{will92Ay} in the choice of error measures, which are the relative $L_2$- and $L_\infty$-errors, ${\rm RE}_2(t)$ and ${\rm RE}_\infty(t)$ at time $t$, with the $L_2$-error being weighted by the cosine of the latitude, which we approximate as follows
\begin{align*}
{\rm RE}_2^h(t) &= \frac{\left(\sum\limits_{i=1}^{N}\big(h^{\boldsymbol{\theta}}(t,\lambda^i,\theta^i)-h(t,\lambda^i,\theta^i)\big)^2\cos\theta^i\right)^{1/2}}{\left(\sum\limits_{i=1}^{N}h(t,\lambda^i,\theta^i)^2\cos\theta^i\right)^{1/2}},\\[1ex]
{\rm RE}_\infty^h(t) &= \frac{\max_{i\in\{1,\dots,N\}}\big|h^{\boldsymbol{\theta}}(t,\lambda^i,\theta^i)-h(t,\lambda^i,\theta^i)\big|}{\max_{i\in\{1,\dots,N\}}\big|h(t,\lambda^i,\theta^i)\big|},
\end{align*}
and
\begin{align*}
{\rm RE}_2^\mathbf{v}(t) &= \frac{\left(\sum\limits_{i=1}^{N}\Big(\big(u^{\boldsymbol{\theta}}(t,\lambda^i,\theta^i)-u(t,\lambda^i,\theta^i)\big)^2+ \big(v^{\boldsymbol{\theta}}(t,\lambda^i,\theta^i)-v(t,\lambda^i,\theta^i)\big)^2\Big)\cos\theta^i\right)^{1/2}}{\left(\sum\limits_{i=1}^{N}h(t,\lambda^i,\theta^i)^2\cos\theta^i\right)^{1/2}},\\
{\rm RE}_\infty^\mathbf{v}(t) &= \frac{\max_{i\in\{1,\dots,N\}}\Big(\big(u^{\boldsymbol{\theta}}(t,\lambda^i,\theta^i)-u(t,\lambda^i,\theta^i)\big)^2+ \big(v^{\boldsymbol{\theta}}(t,\lambda^i,\theta^i)-v(t,\lambda^i,\theta^i)\big)^2\Big)^{1/2}}{\max_{i\in\{1,\dots,N\}}\big(u(t,\lambda^i,\theta^i)^2 + v(t,\lambda^i,\theta^i)^2\big)^{1/2}}.
\end{align*}
Here, $N$ denotes the total number of grid points being used for approximating the errors. Below we use a total of $N=N_{\rm lon}\times N_{\rm lat}$ grid points where $N_{\rm lon}=150$ and $N_{\rm lat}=75$ in the longitude and latitude directions, respectively.

The results of these hyperparameter studies are presented in Table~\ref{tab:Williamson2x3}, which contain the means of the above errors over several independent runs using the given network architectures with different random initial network weights.

\begin{table}[!ht]
\setlength\abovecaptionskip{.3\baselineskip}
\centering
\caption{Convergence study for the nonlinear zonal geostrophic flow test case using three neural networks. Reported are the relative errors on the final time $t=5$ days.}%
\label{tab:Williamson2x3}
\begin{tabular}{lccc}
\hline
$N_\Delta/N_{\rm i}$ & $10^3 / 10^2$ & $10^4 / 10^3$ & $10^5 / 10^4$ \\
\hline
\multicolumn{4}{c}{$4$ layers and $20$ units per layer}\\
\hline
${\rm RE}_2^h$ & $4.59\cdot 10^{-3}$& $1.51\cdot 10^{-3}$ & $5.02\cdot 10^{-4}$  \\
${\rm RE}_\infty^h$ &$2.13\cdot 10^{-2}$ & $1.02\cdot 10^{-2}$ & $3.21\cdot 10^{-3}$ \\
${\rm RE}_2^\mathbf{v}$ & $2.61\cdot 10^{-2}$& $8.67\cdot 10^{-3}$ & $2.82\cdot 10^{-3}$\\
${\rm RE}_\infty^\mathbf{v}$ &$7.72\cdot 10^{-2}$ & $4.20\cdot 10^{-2}$ & $2.28\cdot 10^{-2}$\\
\hline
\multicolumn{4}{c}{$6$ layers and $60$ units per layer}\\
\hline
${\rm RE}_2^h$ & $5.33\cdot 10^{-3}$& $2.34\cdot 10^{-3}$ & $4.85\cdot 10^{-4}$\\
${\rm RE}_\infty^h$ &$2.68\cdot 10^{-2}$ & $7.40\cdot 10^{-3}$ & $2.19\cdot 10^{-3}$\\
${\rm RE}_2^\mathbf{v}$ & $2.52\cdot 10^{-2}$& $9.28\cdot 10^{-3}$ & $3.35\cdot 10^{-3}$ \\
${\rm RE}_\infty^\mathbf{v}$ &$8.40\cdot 10^{-2}$ & $3.49\cdot 10^{-2}$ & $2.29\cdot 10^{-2}$\\
\hline
\multicolumn{4}{c}{$8$ layers and $80$ units per layer}\\
\hline
${\rm RE}_2^h$ & $4.39\cdot 10^{-3}$& $7.32\cdot 10^{-4}$ & $2.07\cdot 10^{-4}$\\
${\rm RE}_\infty^h$ &$1.82\cdot 10^{-2}$ & $6.24\cdot 10^{-3}$ & $1.82\cdot 10^{-3}$\\
${\rm RE}_2^\mathbf{v}$ & $2.21\cdot 10^{-2}$& $5.12\cdot 10^{-3}$ & $1.49\cdot 10^{-3}$ \\
${\rm RE}_\infty^\mathbf{v}$ &$6.35\cdot 10^{-2}$ & $3.41\cdot 10^{-2}$ & $1.62\cdot 10^{-2}$\\
\hline
\end{tabular}
\end{table}

The results of these studies confirm that the more collocation points are being used, the more accurate the numerical results become.

\subsection{Zonal flow over an isolated mountain}

The initial conditions for this test case are the same as for the previous test case reported in Section~\ref{sec:NonlinearGeostrophicFlow}, setting $\alpha=0$, $h_0=5960\ {\rm m}$ and $u_0=20\ {\rm m}{\cdot}{\rm s}^{-1}$. In addition, a bottom topography is added in the form of an isolated mountain, the height of which is given by
\[
b = b_0\left(1-\frac{r}{R}\right),
\]
with $b_0=2000\ {\rm m}$, $R=\pi/9$, and $r^2=\min[R^2, (\lambda-\lambda_{\rm c})^2+(\theta-\theta_{\rm c})^2]$, and the center of the mountain being located at $\lambda_{\rm c} = -\pi/2$ and $\theta_{\rm c}=\pi/6$. There is no analytical solution known for this test case, but contour plots of this test case should show the formation of a wave train in the lee (east) of the mountain, which will start to travel through both hemispheres the longer the simulation is ongoing.

We carry out three experiments, differing in the number of collocation points, batch sizes and number of models being used. The architecture for all tests is given by a neural network with 8 layers and 80 units per layer. For the first two experiments we split the total duration of the integration, spanning from day 0 to day 15 into 5 sub-intervals of 3 days each. The learning rates of the 5 separate models were $10^{-3}$, $10^{-3}/\sqrt{10}$, $10^{-4}$, $10^{-4}$ and $10^{-4}$ respectively. Each model is being trained for $15000$ epochs, that is, the total training spans $75000$ epochs. The last experiment uses 6 sub-intervals of 2.5 days each, and the learning rates of the 6 models were $10^{-3}$, $10^{-3}/\sqrt{10}$, $10^{-3}/\sqrt{10}$, $10^{-4}$, $10^{-4}$ and $10^{-4}$ respectively. Also here each model is being trained for $15000$ epochs, therefore the total training spans $90000$ epochs. Note that we progressively reduce the learning rates for each sub-model except for the first one as we are using the weights of the previous model as the initial weights for the next model. This endows the multi-model approach to training physics-informed neural networks with aspects of transfer learning, with the assumption that a trained sub-model provides a good starting point for the yet-to-be-trained next sub-model, thus justifying the use of smaller learning rates.

For the first experiment, we use $N_{\Delta}=10^5$ PDE collocation points and $N_{0}=10^4$ initial collocation points with batch size 500 each for computing $\mathcal L_{\Delta}$ and $\mathcal L_{\rm i}$. For the second experiment we keep the number of PDE and initial collocation points as in the first experiment, but reduce the batch sizes to $125$ for computing both $\mathcal L_{\Delta}$ and $\mathcal L_{\rm i}$. For the third experiment we set $N_{\Delta}=9\cdot 10^5$ and $N_{0}=3\cdot10^4$, with batch sizes $1875$ and $375$, respectively, for computing $\mathcal L_{\Delta}$ and $\mathcal L_{\rm i}$.

\begin{figure}[!ht]
\centering
\begin{subfigure}{0.33\textwidth}
  \centering
  \includegraphics[width=\linewidth]{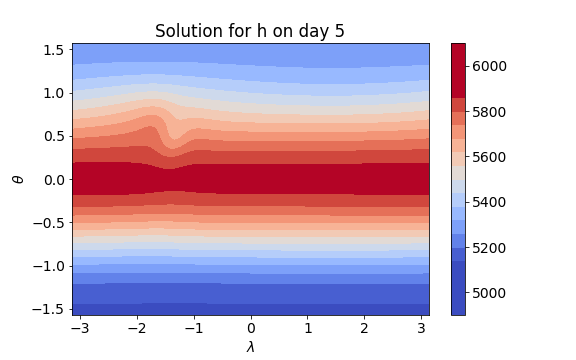}
\end{subfigure}
\begin{subfigure}{0.33\textwidth}
  \centering
  \includegraphics[width=\linewidth]{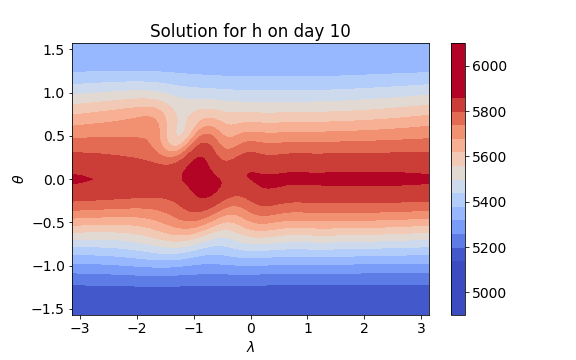}
\end{subfigure}
\begin{subfigure}{0.32\textwidth}
  \centering
  \includegraphics[width=\linewidth]{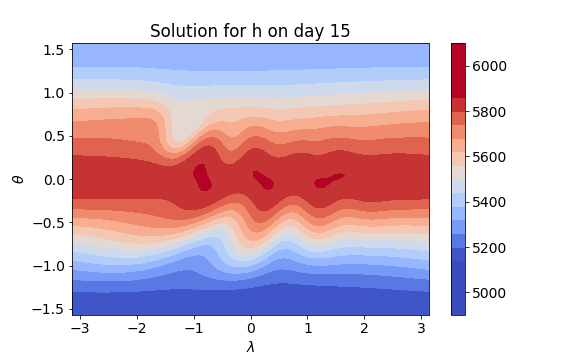}
\end{subfigure}
\begin{subfigure}{0.33\textwidth}
  \centering
  \includegraphics[width=\linewidth]{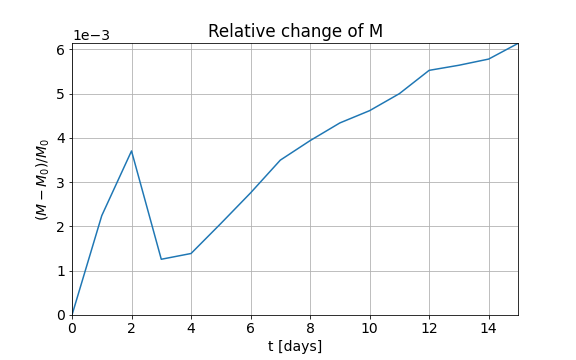}
\end{subfigure}
\begin{subfigure}{0.33\textwidth}
  \centering
  \includegraphics[width=\linewidth]{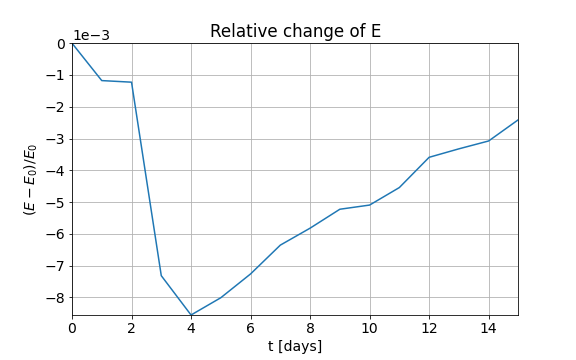}
\end{subfigure}
\begin{subfigure}{0.32\textwidth}
  \centering
  \includegraphics[width=\linewidth]{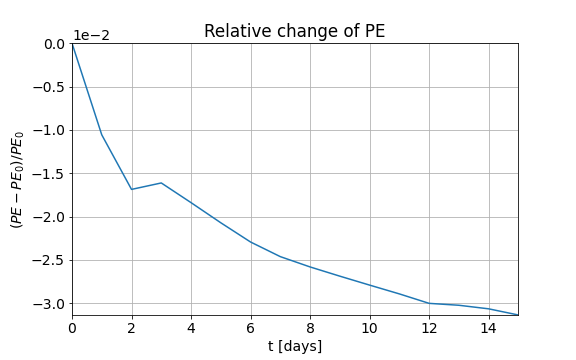}
\end{subfigure}
\\[3ex]
\begin{subfigure}{0.33\textwidth}
  \centering
  \includegraphics[width=\linewidth]{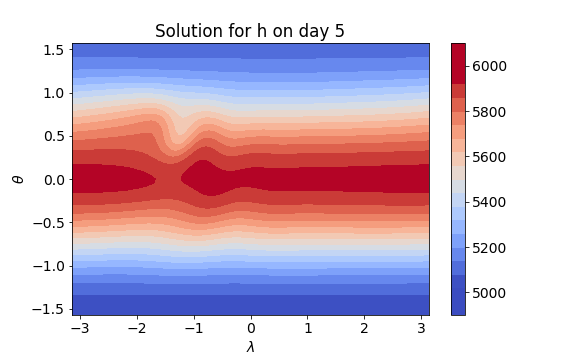}
\end{subfigure}
\begin{subfigure}{0.33\textwidth}
  \centering
  \includegraphics[width=\linewidth]{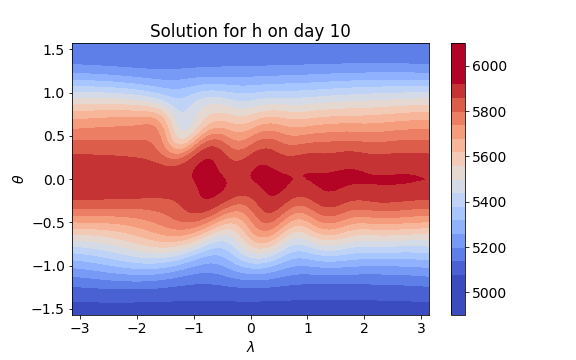}
\end{subfigure}
\begin{subfigure}{0.32\textwidth}
  \centering
  \includegraphics[width=\linewidth]{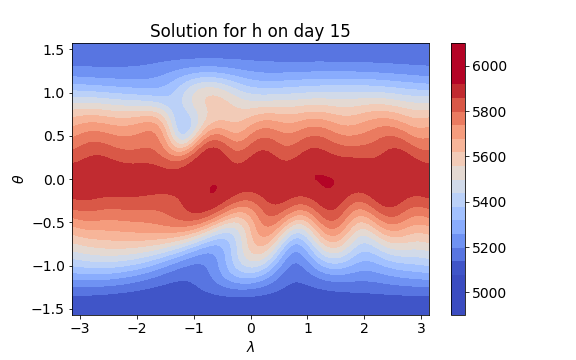}
\end{subfigure}
\begin{subfigure}{0.33\textwidth}
  \centering
  \includegraphics[width=\linewidth]{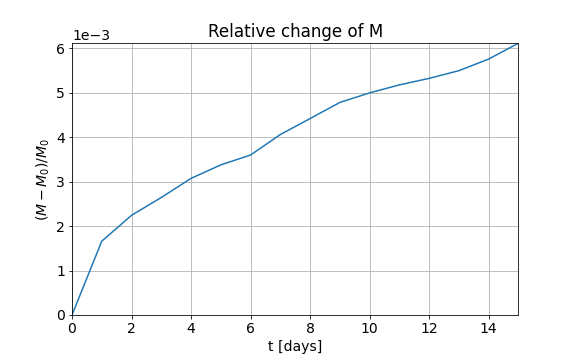}
\end{subfigure}
\begin{subfigure}{0.33\textwidth}
  \centering
  \includegraphics[width=\linewidth]{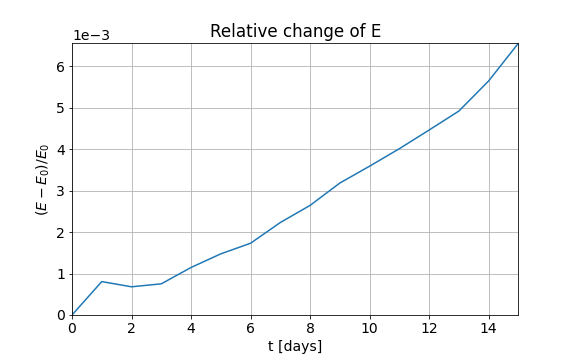}
\end{subfigure}
\begin{subfigure}{0.32\textwidth}
  \centering
  \includegraphics[width=\linewidth]{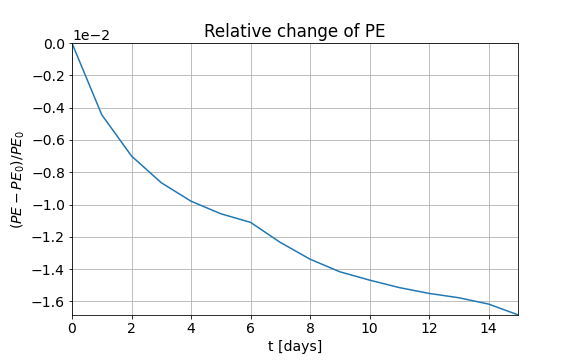}
\end{subfigure}
\\[3ex]
\begin{subfigure}{0.33\textwidth}
  \centering
  \includegraphics[width=\linewidth]{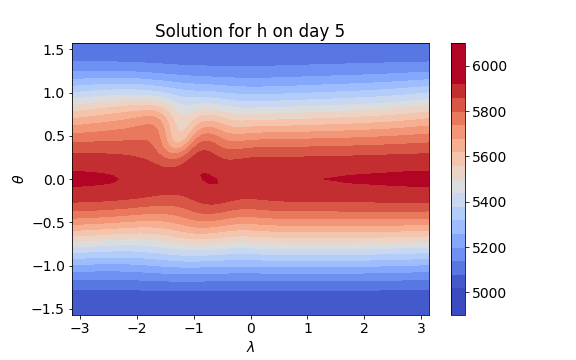}
\end{subfigure}
\begin{subfigure}{0.33\textwidth}
  \centering
  \includegraphics[width=\linewidth]{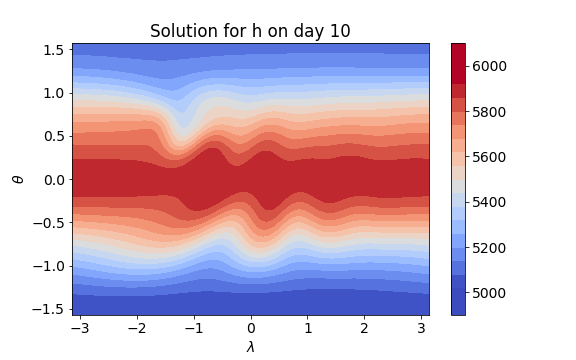}
\end{subfigure}
\begin{subfigure}{0.32\textwidth}
  \centering
  \includegraphics[width=\linewidth]{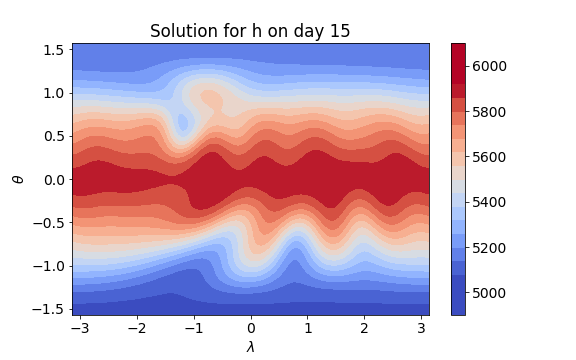}
\end{subfigure}
\begin{subfigure}{0.33\textwidth}
  \centering
  \includegraphics[width=\linewidth]{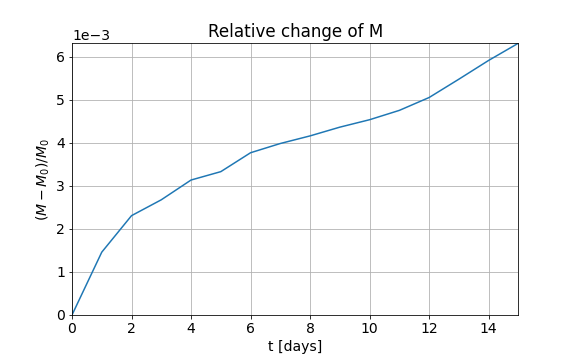}
\end{subfigure}
\begin{subfigure}{0.33\textwidth}
  \centering
  \includegraphics[width=\linewidth]{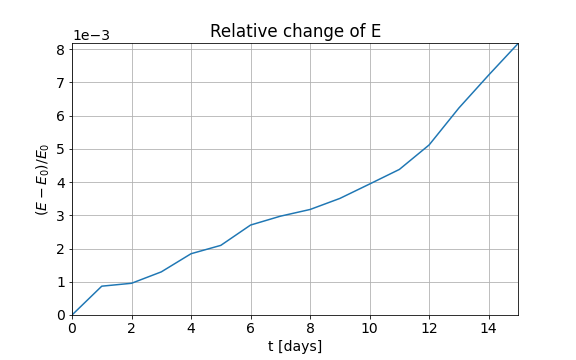}
\end{subfigure}
\begin{subfigure}{0.32\textwidth}
  \centering
  \includegraphics[width=\linewidth]{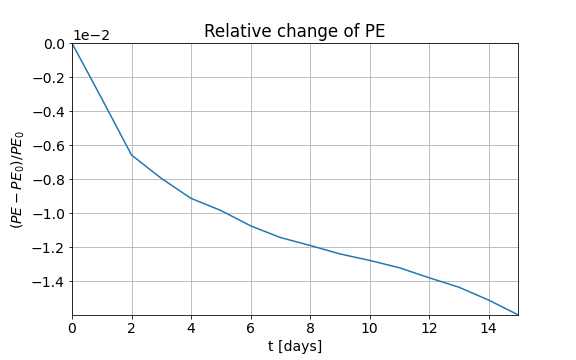}
\end{subfigure}
\\[2ex]${ }$
\caption{Results on the flow over an isolated mountain test case proposed in~\cite{will92Ay}. \textit{Top two rows:} $N_\Delta=10^5$, $N_{\rm i}=10^4$ with batch sizes $500$ each for computing $\mathcal L_{\Delta}$ and $\mathcal L_{\rm i}$, respectively. \textit{Middle two rows:} $N_\Delta=10^5$, $N_{\rm i}=10^4$ with batch sizes $125$ each for computing $\mathcal L_{\Delta}$ and $\mathcal L_{\rm i}$, respectively. \textit{Bottom two rows:} $N_\Delta=9\cdot10^5$, $n_{\rm i}=3\cdot10^4$ with batch sizes $1875$ and $375$ for computing $\mathcal L_{\Delta}$ and $\mathcal L_{\rm i}$, respectively. Depicted are the neural network solutions at days $5$, $10$ and $15$ for the height field $h$, along with the time series of relative change in mass, energy and potential enstrophy.}\label{fig:FlowOverMountain}\vspace{-5ex}
\end{figure}

\begin{figure}[!ht]
\centering
\begin{subfigure}{0.33\textwidth}
  \centering
  \includegraphics[width=\linewidth]{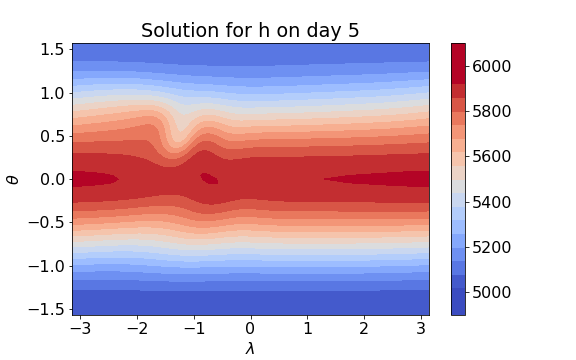}
\end{subfigure}
\begin{subfigure}{0.33\textwidth}
  \centering
  \includegraphics[width=\linewidth]{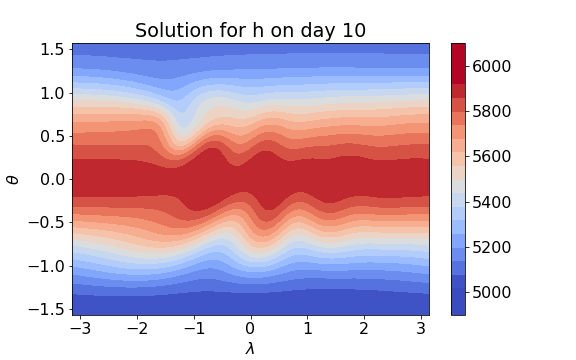}
\end{subfigure}
\begin{subfigure}{0.32\textwidth}
  \centering
  \includegraphics[width=\linewidth]{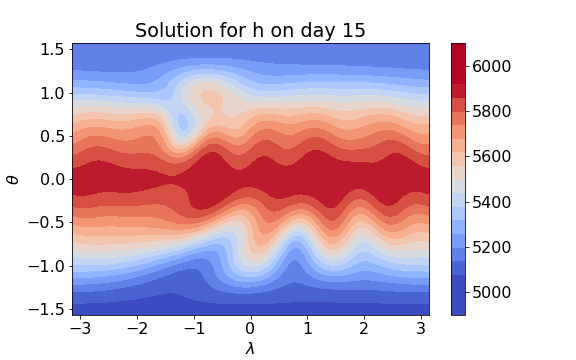}
\end{subfigure}
\\[3ex]
\begin{subfigure}{0.33\textwidth}
  \centering
  \includegraphics[width=\linewidth]{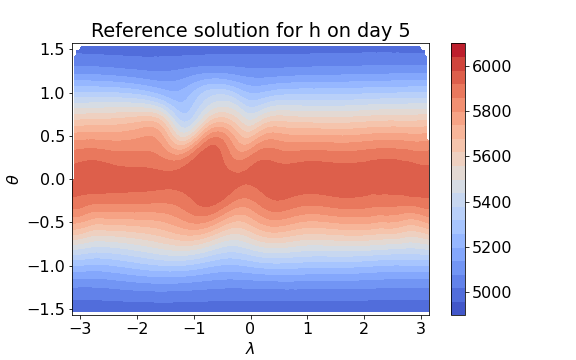}
\end{subfigure}
\begin{subfigure}{0.33\textwidth}
  \centering
  \includegraphics[width=\linewidth]{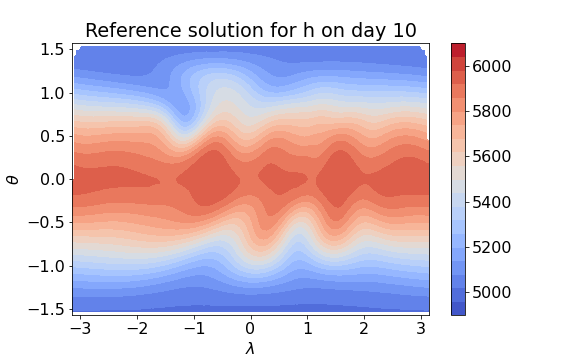}
\end{subfigure}
\begin{subfigure}{0.32\textwidth}
  \centering
  \includegraphics[width=\linewidth]{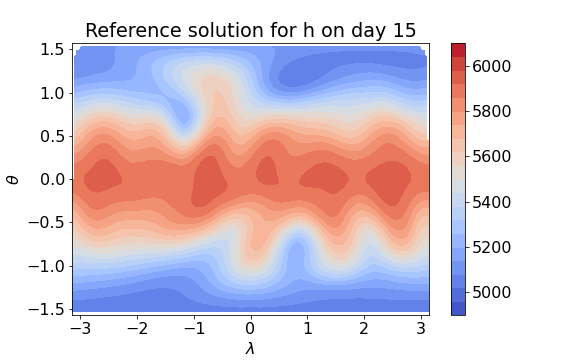}
\end{subfigure}
\\[3ex]
\begin{subfigure}{0.33\textwidth}
  \centering
  \includegraphics[width=\linewidth]{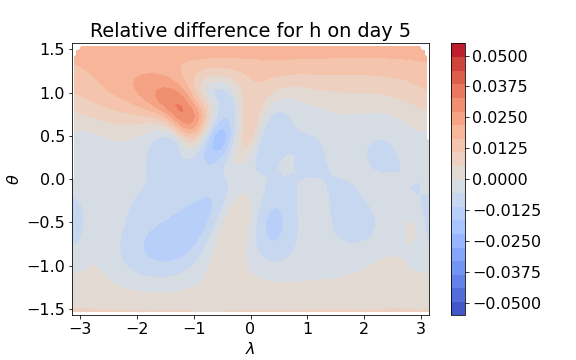}
\end{subfigure}
\begin{subfigure}{0.33\textwidth}
  \centering
  \includegraphics[width=\linewidth]{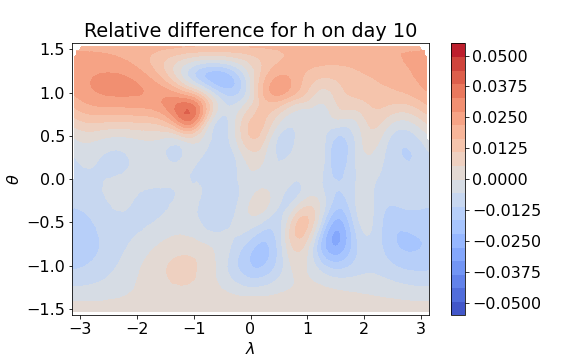}
\end{subfigure}
\begin{subfigure}{0.32\textwidth}
  \centering
  \includegraphics[width=\linewidth]{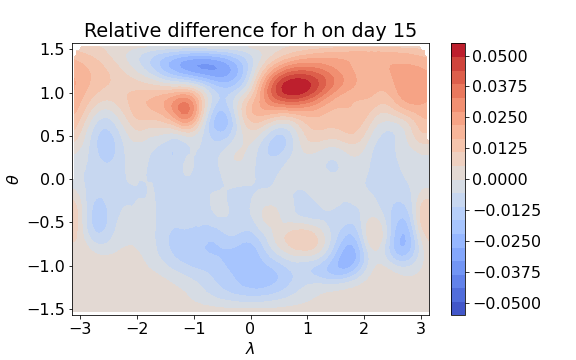}
\end{subfigure}
\\[2ex]${ }$
\caption{Comparison results on the flow over an isolated mountain test case proposed in~\cite{will92Ay}. \textit{Top row:} $N_\Delta=9\cdot10^5$, $n_{\rm i}=3\cdot10^4$ with batch sizes $1875$ and $375$ for computing $\mathcal L_{\Delta}$ and $\mathcal L_{\rm i}$, respectively. \textit{Middle row:} Reference solution using the variational integrator proposed in~\cite{brec19a}. \textit{Bottom row:} Relative difference between the physics-informed neural network solution and the variational integrator.
Depicted are the solutions for the height field $h$ at days $5$, $10$ and $15$ for the height field $h$.}\label{fig:FlowOverMountainComparison}
\end{figure}

The contour plots of the solution at days $5$, $10$ and $15$ along with the time series of mass, energy and potential enstrophy are depicted in Figure~\ref{fig:FlowOverMountain}. These three conserved quantities are respectively given by
\[
\mathcal M = \int_S h^*\mathrm{d}S,\qquad \mathcal E=\frac{1}{2}\int_S h^*\mathbf{v}^2+g(h^2-b^2)\mathrm{d}S,\qquad \mathcal {P} = \frac12\int_S\frac{(\zeta+f)^2}{h^*}\mathrm{d}S,
\]
where $S=[-\pi,\pi]\times[-\pi/2,\pi/2]$, and $\zeta$ is the relative vorticity, $\zeta=(a\cos\theta)^{-1}\big(v_\lambda-(u\cos\theta)_\theta\big)$. Comparison with the same plots given in, e.g.,~\cite{brec19a,jako95a} shows a good visual alignment of the results for this test case, especially for the experiments with smaller batch size and/or using more collocation points. Both these experiments are able to capture the trough of the wave after the mountain more accurately, while the experiment with the fewest collocation points and largest batch sizes tends to be overly diffusive, in particular in regards to the number of waves that are formed in the lee of the mountain. The comparison to the variational integrator developed in~\cite{brec19a} is depicted in Fig.~\ref{fig:FlowOverMountainComparison}.

We should like to point out that all three experiments are relatively coarse-scale, in particular in comparison to traditional numerical simulations. Thus, we use 100000 collocation points. For better reference, assume a constant grid spacing in both time and space, the same number of grid points corresponds roughly to a grid of 46 points each in latitude, longitude and the temporal direction. This is only a negligibly small amount of the spatio-temporal grid points being used in traditional numerical methods. In other words, the results reported for this and the other test cases show how accurate a solution is achievable using physics-informed neural networks for a really small number of collocation points.

\subsection[Rossby–Haurwitz waves]{Rossby--Haurwitz waves}
\label{sec:RossbyHaurwitzWave}

The initial conditions for this test case are
\begin{align*}
 u\big|_{t=0}&=aK \cos \theta + aK \cos^{\kappa-1}\theta (\kappa \sin^2\theta -\cos^2\theta) \cos\kappa\lambda,\\[0.5ex]
 v\big|_{t=0}&=-aK\kappa\cos^{\kappa-1}\theta\sin\theta \sin\kappa\lambda,\\[0.5ex]
 gh\big|_{t=0}&=gh_0+a^2A(\theta)+a^2B(\theta)\cos\kappa\lambda + a^2 C(\theta) \cos 2\kappa\lambda,
\end{align*}
where
\begin{align*}
A(\theta)&=\tfrac12 K(2\omega+K)\cos^2\theta +\tfrac 1 4 K^2 \cos^{2\kappa} \theta \Big( (\kappa+1)\cos^2\theta + (2\kappa^2-\kappa-2)-2\kappa^2\cos^{-2}\theta \Big),\\[1.5ex]
B(\theta)&=\frac{2(\omega+K)K}{(\kappa+1)(\kappa +2)}\cos^{\kappa}\theta \Big( (\kappa^2+2\kappa+2)-(\kappa+1)^2\cos^2 \theta \Big),\\[1.5ex]
C(\theta)&= \tfrac{1}{4} K^2 \cos^{2\kappa}\theta \Big( (\kappa+1)\cos^2\theta-(\kappa+2) \Big).
\end{align*}		
The parameters used are $K=7.848\cdot 10^{-6}\ {\rm s}^{-1}$, $\kappa= 4$ and $h_0= 8000\ {\rm m}$. Note that while Rossby--Haurwitz waves are not exact solutions to the shallow-water equations on the sphere they are still a standard test case for this model, which should be capable of simulating the propagation of these important waves.

We split the time interval of integration $[0,t_{\rm f}]$ with $t_{\rm f}=14$ days into five equally long subintervals. Each subinterval model is a neural network with 8 layers and 80 units per layer. The first model is trained for 40000 epochs, all other models for 20000 epochs, that is, the total training spans 120000 epochs. The learning rate for the first model is $\eta_1=10^{-3}$, the learning rate for the second model is $\eta_2=10^{-3}/\sqrt{10}$, and the learning rates for the remaining models are $\eta_{3,4,5}=10^{-4}$. We use $N_\Delta=10^5$ collocation points over the entire spatio-temporal domain~$\Omega$ and $N_{\rm i}=10^4$ initial points.

\begin{figure}[!t]
\centering
\begin{subfigure}{0.33\textwidth}
  \centering
  \includegraphics[width=\linewidth]{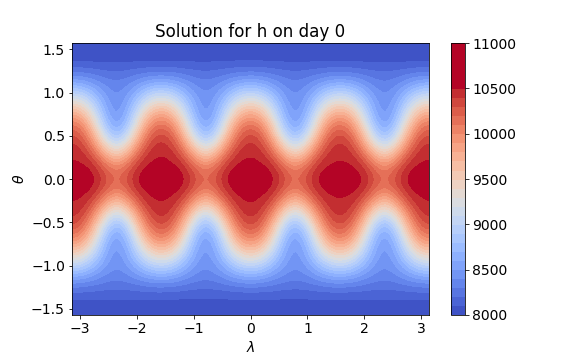}
\end{subfigure}
\begin{subfigure}{0.33\textwidth}
  \centering
  \includegraphics[width=\linewidth]{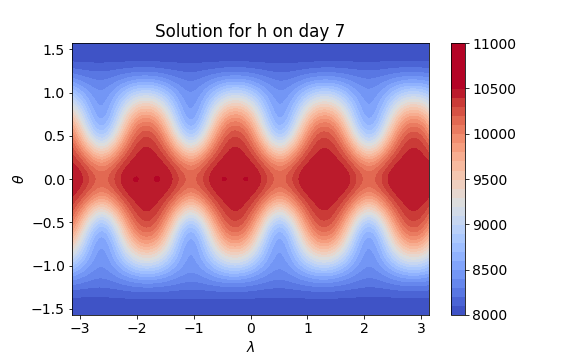}
\end{subfigure}
\begin{subfigure}{0.32\textwidth}
  \centering
  \includegraphics[width=\linewidth]{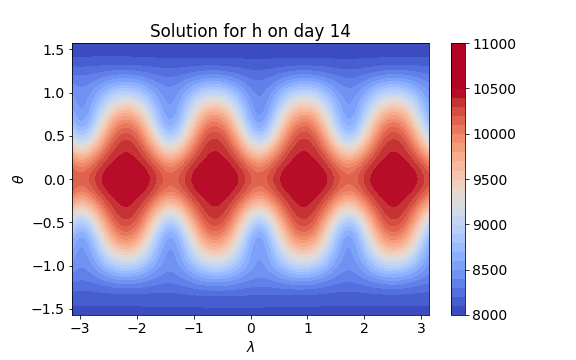}
\end{subfigure}
\begin{subfigure}{0.33\textwidth}
  \centering
  \includegraphics[width=\linewidth]{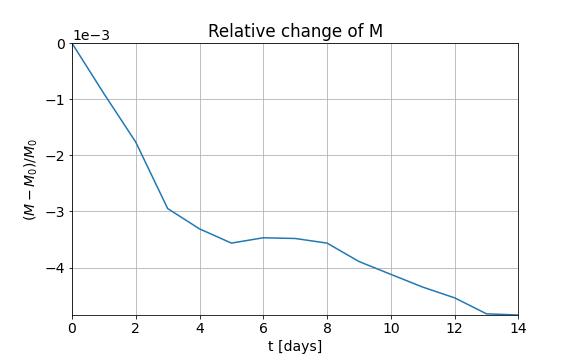}
\end{subfigure}
\begin{subfigure}{0.33\textwidth}
  \centering
  \includegraphics[width=\linewidth]{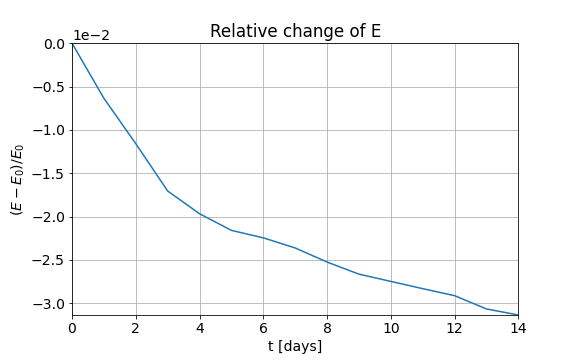}
\end{subfigure}
\begin{subfigure}{0.32\textwidth}
  \centering
  \includegraphics[width=\linewidth]{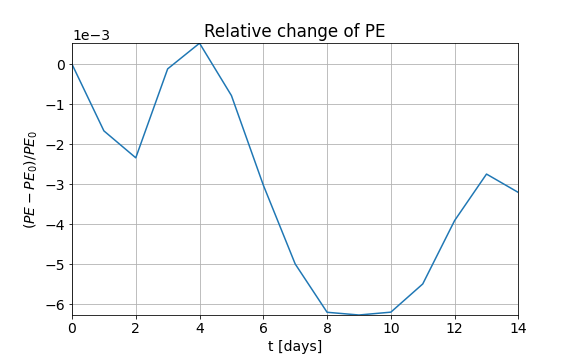}
\end{subfigure}
\caption{Results on the Rossby--Haurwitz wave test case proposed in~\cite{will92Ay}. \textit{Top row:} Solution for the height field $h$ at days $0$, $7$, and $14$. \textit{Bottom row:} Relative change of mass, energy and potential enstrophy.}\vspace{1.ex}
\label{fig:WilliamsonTest6}
\end{figure}

The results of this experiment are depicted in Figure~\ref{fig:WilliamsonTest6}.

\begin{figure}[!t]
\centering
\begin{subfigure}{0.33\textwidth}
  \centering
  \includegraphics[width=\linewidth]{RHWave_day_0.png}
\end{subfigure}
\begin{subfigure}{0.33\textwidth}
  \centering
  \includegraphics[width=\linewidth]{RHWave_day_7.png}
\end{subfigure}
\begin{subfigure}{0.32\textwidth}
  \centering
  \includegraphics[width=\linewidth]{RHWave_day_14.png}
\end{subfigure}
\begin{subfigure}{0.33\textwidth}
  \centering
  \includegraphics[width=\linewidth]{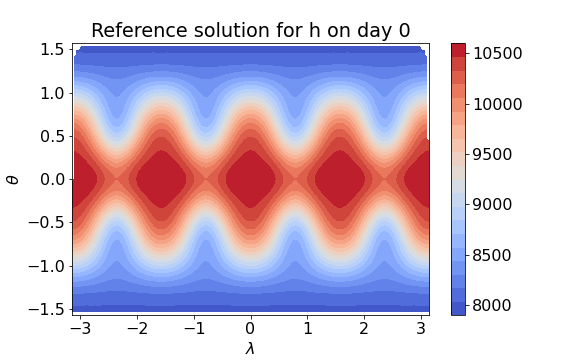}
\end{subfigure}
\begin{subfigure}{0.33\textwidth}
  \centering
  \includegraphics[width=\linewidth]{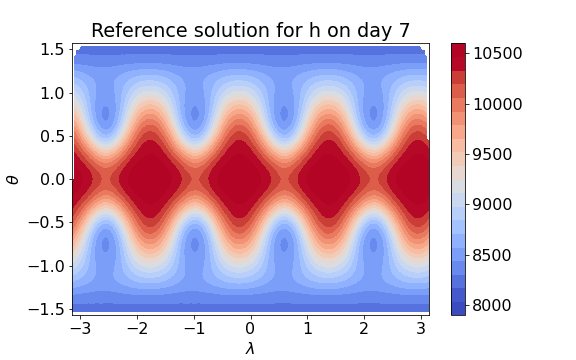}
\end{subfigure}
\begin{subfigure}{0.32\textwidth}
  \centering
  \includegraphics[width=\linewidth]{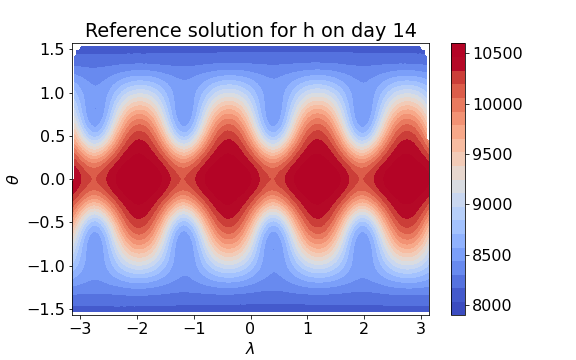}
\end{subfigure}
\begin{subfigure}{0.33\textwidth}
  \centering
  \includegraphics[width=\linewidth]{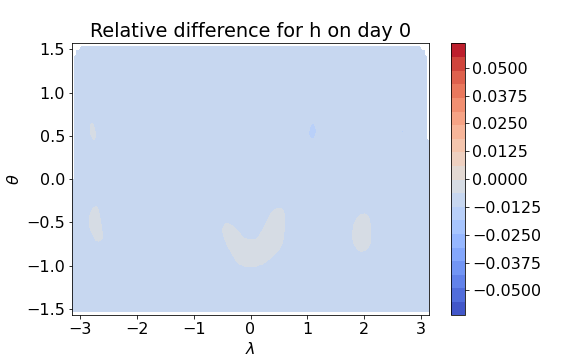}
\end{subfigure}
\begin{subfigure}{0.33\textwidth}
  \centering
  \includegraphics[width=\linewidth]{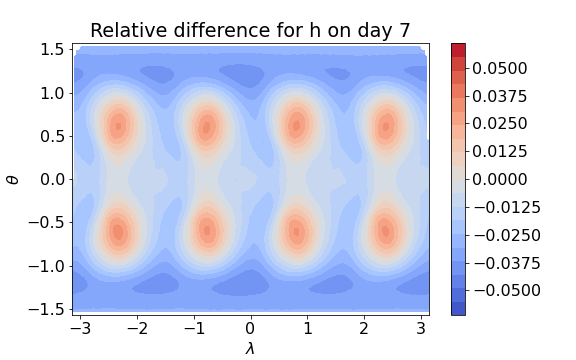}
\end{subfigure}
\begin{subfigure}{0.32\textwidth}
  \centering
  \includegraphics[width=\linewidth]{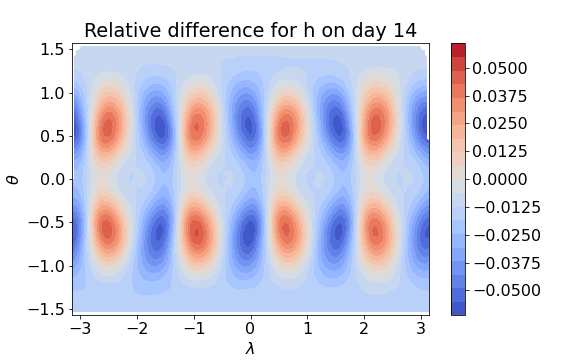}
\end{subfigure}
\caption{Results on the Rossby--Haurwitz wave test case proposed in~\cite{will92Ay}. \textit{Top row:} Solution for the height field $h$ at days $0$, $7$, and $14$. \textit{Middle row:} Reference solution using the variational integrator proposed in~\cite{brec19a}. \textit{Bottom row}: Relative difference between the physics-informed neural network solution and the variational integrator. Depicted are the solutions at days 0, 7 and 14 for the height field $h$.}\vspace{1.5ex}
\label{fig:WilliamsonTest6Comparison}
\end{figure}

As for the previous test case, also the results reported are in good visual agreement with the results presented in~\cite{brec19a,jako95a}, showing that physics-informed neural networks are indeed able to simulate the propagation of Rossby--Haurwitz waves in the shallow-water model. A quantitative comparison against the solution obtained using the variational integrator developed in~\cite{brec19a} is shown in Fig.~\ref{fig:WilliamsonTest6Comparison}.

\section{Conclusion}\label{sec:ConclusionsPINNsSWE}

In this paper we have developed physics-informed neural networks for solving the shallow-water equations on the sphere. While we found that the vanilla algorithm as proposed in~\cite{rais19a} did not allow us to train the required neural networks for the time horizons of interest for the benchmarks proposed in~\cite{will92Ay}, two straightforward variations of the basic physics-informed neural network training algorithm alleviated these issues. We firstly split the entire time interval $[0,t_{\rm f}]$ into many non-overlapping subintervals and successively train neural networks for each subinterval, using the predictions for the final time of the previous network as the initial condition for the next network. By incorporating the periodic boundary conditions as hard constraints into the neural network architectures we avoid having to include them as additional term in the loss function. Therefore, the loss function to be minimized is the sum of only the PDE loss and the initial value loss, which significantly simplifies the optimization procedure. To mitigate the conflicting gradient problem of the multi-task learning problem associated with the neural networks having to learn to satisfy the initial conditions and the system of partial differential equations at the same time we used a projected-gradient approach which will correct conflicting gradients and thus making sure that the gradient descent algorithm will not stall at a minimum related to only one part of the composite loss function.

The shallow-water equations are typically used as a test bed in numerical meteorology for proposing new algorithms for eventually solving the full set of governing hydro-thermodynamical equations governing the evolution of the atmosphere--ocean system. As such, the eventual application of physics-informed neural networks for the same purpose is the main motivation for the present work.

A few issues will have to be addressed for the problem of weather forecasting to be tackled with physics-informed neural networks. There are several essential and desirable properties a dynamical core of a weather forecasting system should possess, including \textit{mass conservation}, \textit{energy conservation}, an \textit{accurate representation of balanced flow and adjustment}, and \textit{minimal grid imprinting, see~\cite{stan12a} for a full list of these properties along with some discussion}.

As evidenced by the presented results, physics-informed neural networks are not automatically conservative, although the issue of conservation has recently been addressed in~\cite{jagt20a} via domain decomposition and a straightforward application of the divergence theorem. Thus, while more work may be needed in this regard, some potentially feasible ways to incorporate conservation into physics-informed neural networks have already been proposed. We intend to investigate the possibility of including conservation laws as hard constraints in physics-informed neural networks in the future. Regarding the issue of grid imprinting, this is where physics-informed neural networks should have a clear advantage over standard grid-based numerical methods, as no fixed meshes have to be used when training them. That is, physics-informed neural networks are by definition \textit{meshless}, and once they have been trained they can be evaluated at any point within their spatio-temporal domain of definition. Ensuring the validity of the other properties discussed in~\cite{stan12a} will have to be the subject of more in-depth investigations.

A further challenge for the physics-informed neural network paradigm to be used for weather and climate prediction is that realistic forecast models have to include physical parameterization schemes~\cite{sten07a}, that on a practical level may not be able to be written in the form of a system of differential equations. In this case it may still be possible to use a neural-network-based solution algorithm, albeit for the discretized and parameterized form of the system of equations to be solved. For example, in~\cite{rais19a} a discrete time version of physics-informed neural networks was developed that can serve as an alternative to high-order Runge--Kutta methods. A more in-depth investigation of a discretized version of neural-network-based solvers for geophysical fluid mechanics will have to be carried out to assess whether the proposed methodology can be used for meteorological models that include physical parameterization schemes.

Lastly, training times of physics-informed neural networks for the governing equations of hydro-thermodynamics will be critical as well. Traditional numerical weather predictions require on the time-scale of a few hours to produce multi-day forecasts of all meteorological fields, with roughly 2 hours of this time being devoted to data assimilation and 1--2 hours being required for the numerical integration of the governing equations~\cite{haim21a}. Since physics-informed neural networks would provide an alternative to traditional numerical methods for the governing equations of a weather forecasting system, training times would have to be competitively fast to allow for issuing weather forecasts in real-time. The training times we observed while carrying out the present study are roughly in line with the computational time being used in traditional numerical weather prediction, although a further reduction of the physics-informed neural network training times may be possible using multi-GPU systems. This has for example been demonstrated in~\cite{shuk21a}. In turn, the multi-model approach put forth in the present paper has by nature a serial bottleneck in that subsequent models have to be trained subsequently. That is, each model working on a spatio-temporal slice has to complete training before the next model working on the subsequent spatio-temporal slice can begin training. We are currently in the process to investigate how this serial bottleneck can be alleviated, for example using a parallel-in-time version of physics-informed neural networks as proposed in~\cite{meng20a}. Despite this bottleneck, due to the formulation of the multi-model approach presented here, results of the numerical solution will become progressively available from the initial time $t=0$ the longer the training goes on. That is, it is not necessary to wait for the completion of the algorithm to get an increasing part of the numerical solution.

We should also highlight again that the computational results obtained by using physics-informed neural networks capture the main properties of the standard test cases of~\cite{will92Ay} albeit using only a fraction of the computational nodes that have to be used in standard numerical methods. Scaling up the number of collocation points and the capacity of the neural networks being used is expected to improve results even further, as was corroborated in our convergence studies.
\looseness=-1

\section*{Acknowledgements}

The authors are grateful to the anonymous referees for a number of valuable remarks and suggestions.
The authors also thank Leopold Haimberger for helpful discussions and R\"udiger Brecht for providing the numerical results of the variational integrators for comparison for two test cases.
This research was undertaken thanks to funding from the Canada Research Chairs program, the InnovateNL LeverageR{\&}D program and the NSERC Discovery Grant program. The research of ROP was supported by the Austrian Science Fund (FWF), projects P28770 and P30233.
\looseness=-1

\footnotesize

\end{document}